\documentclass[aps, amsfonts, amssymb, amsmath, reprint, showkeys, nofootinbib, superscriptaddress]{revtex4-2}
\usepackage[utf8]{inputenc}

\usepackage{tabularx}
\usepackage[colorinlistoftodos, color=green!40, prependcaption]{todonotes}
\usepackage{dcolumn}
\usepackage{amsthm}
\usepackage{mathtools}
\usepackage{physics}
\usepackage{xcolor}
\usepackage{graphicx}
\usepackage{adjustbox}
\usepackage{placeins}
\usepackage{bm}
\usepackage[T1]{fontenc}
\usepackage{subfigure}
\usepackage[version=4]{mhchem}
\usepackage{array}
\usepackage{booktabs} 
\renewcommand{\arraystretch}{1.3} 
\usepackage{multirow}
\usepackage{csquotes}
\usepackage{siunitx}
\usepackage{comment}
\usepackage[pdftex, pdftitle={Article}, pdfauthor={Author}, colorlinks=true]{hyperref}
\begin{document}
\title{Diamagnetic microchip traps for levitated nanoparticle entanglement experiments}
\author{Shafaq Gulzar Elahi}
    \affiliation{Center for Fundamental Physics, Department of Physics and Astronomy, Northwestern University, 2145 Sheridan Road, Evanston, IL}
\author{Martine Schut}
    \affiliation{Van Swinderen Institute for Particle Physics and Gravity, University of Groningen, 9747AG Groningen, the Netherlands }
    \affiliation{Bernoulli Institute for Mathematics, Computer Science and Artificial Intelligence, University of Groningen, 9747 AG Groningen, the Netherlands \vspace{1mm}}
\author{Andrew Dana}
    \affiliation{Center for Fundamental Physics, Department of Physics and Astronomy, Northwestern University, 2145 Sheridan Road, Evanston, IL}
\author{Alexey Grinin}
    \affiliation{Center for Fundamental Physics, Department of Physics and Astronomy, Northwestern University, 2145 Sheridan Road, Evanston, IL}
      
  \author{Sougato Bose}
    \affiliation{Department of Physics and Astronomy, University College London, London WC1E 6BT, United Kingdom}
\author{Anupam Mazumdar }
    \affiliation{Van Swinderen Institute for Particle Physics and Gravity, University of Groningen, 9747AG Groningen, the Netherlands }
 \author{Andrew Geraci}
    \affiliation{Center for Fundamental Physics, Department of Physics and Astronomy, Northwestern University, 2145 Sheridan Road, Evanston, IL}

\begin{abstract}
The Quantum Gravity Mediated Entanglement (QGEM) protocol offers a novel method to probe the quantumness of gravitational interactions at non-relativistic scales. This protocol leverages the Stern-Gerlach effect to create $\mathcal{O}(\mu m)$ spatial superpositions of two nanodiamonds (mass $\sim 10^{-15}$ kg) with NV spins, which are then allowed to interact and become entangled solely through the gravitational interaction. Since electromagnetic interactions such as Casimir-Polder and dipole-dipole interactions dominate at this scale, screening them to ensure the masses interact exclusively via gravity is crucial.
In this paper, we propose using magnetic traps based on micro-fabricated wires, which provide strong gradients with relatively modest magnetic fields to trap nanoparticles for interferometric entanglement experiments. The design consists of a small trap to cool the center-of-mass motion of the nanodiamonds and a long trap with a weak direction suitable for creating macroscopic superpositions. 
In contrast to permanent-magnet-based long traps, the micro-fabricated wire-based approach allows fast switching of the magnetic trapping and state manipulation potentials and permits integrated superconducting shielding, which can screen both electrostatic and magnetic interactions between nanodiamonds in a gravitational entanglement experiment. The setup also provides a possible platform for other tests of quantum coherence in macroscopic systems and searches for novel short-range forces.
\end{abstract}


\maketitle

\section{Introduction}\label{sec:intro}

The nature of spacetime or the origin of the gravitational interaction poses a stringent challenge for theoretical physics. We still do not understand the quantum origin of the gravitational interaction. Given the weakness of the interaction, the direct detection of a graviton, quanta similar to a photon in the case of quantum electrodynamics (QED), is impossible~\cite{dyson_is_2013}.

However, a protocol known as the quantum-
gravity-induced entanglement of masses (QGEM)~\cite{Bose:2017nin,ICTS}, see also~\cite{Marletto:2017kzi}, has provided a novel way of testing the quantum features of gravity in a table-top experiment~\cite{Bose:2017nin}. It does so by witnessing the quantum entanglement~\cite{Horodecki:2009zz}, a {\it sole} feature of quantum mechanics inherited in the quantum uncertainty principle. Classical realism cannot mimic entanglement between two quantum systems, e.g. in a quantum spatial superposition.  The most basic principle of quantum gravity is that the gravitational interaction can entangle the two masses through the quantum nature of the graviton, which is impervious to how gravity is quantised at the ultraviolet~\cite{Marshman:2019sne,Bose:2022uxe,Vinckers:2023grv}. In the infrared, where the QGEM experiment will be performed, the graviton description of quantising the low energy fluctuations is ideal; see~\cite {Gupta,Donoghue:1994dn}. 

It has been rigorously demonstrated that the quantum nature of gravity will induce entanglement even at the level of Newtonian potential, where there are no $\hbar$ effects at the lowest order in the expansion of Newton's constant~\cite{Bose:2017nin,Carney_2019,
Carney:2021vvt,
Marshman:2019sne,Bose:2022uxe,Danielson:2021egj,Christodoulou:2022mkf,Elahi:2023ozf,Vinckers:2023grv}. If the matter is in a quantum superposition, and gravity is quantum, so is the change in the gravitational potential; hence, the gravitational interaction Hamiltonian at any order in Newton's constant is an operator-valued entity; therefore, the position and momentum involved in the gravitational Hamiltonian are all operator valued entity acting on the respective states of the matter system~\cite{Bose:2022uxe}. Furthermore, a quantum matter can also entangle to photons, a quantum version of the light bending experiment, which will further ensure the spin-2 nature of the gravitational interaction~\cite{Carney:2021vvt,Biswas:2022qto}.

The QGEM protocol relies on creating the adjacent, macroscopic quantum superposition for {\it electrically neutral} masses and bringing them sufficiently close to each other so that the two systems are entangled {\it solely} via the gravitational interaction. However, realizing this in a laboratory is a challenge in itself. The electromagnetic interaction can dominate over the gravitational one due to potentials such as electric and magnetic dipole~\cite{vandeKamp:2020rqh,Schut2024,Schut:2023eux}, Casimir-Polder~\cite{Casimir:1947kzi,Casimir:1948dh}, and patch potentials~\cite{Speake,Kim:2009mr,Behunin_2012}. Notably, an interesting observation was first made in\cite{vandeKamp:2020rqh} that the electromagnetic interactions can be screened while the gravitational interaction cannot be, leading to various analyses on the designing aspects of the QGEM experiment~
\cite{Schut2024}. We further studied the role of dephasing and decoherence due to Casimir-Polder potential and an electric dipole in the case of creating spatial superposition in a free-falling setup~\cite{Schut:2023tce,Fragolino:2023agd}. However, more crucially, we realised that trapping and shielding the particle simultaneously leads to further improvements in witnessing the entanglement due to quantum gravity between the two masses~\cite{Schut2024}.
In particular, the superposition size required to test the quantum nature of gravity is now reduced by roughly two orders of magnitude for similar mass scale with the insertion of a conducting plate~\cite{Schut2024}.
\textcolor{black}{Although optical trapping of silica nanoparticles has seen several notable advancements \cite{delic2020cooling,tebbenjohanns2021quantum}, nanodiamonds have not yet been optically trapped in high vacuum, and light scattering poses a technical challenge for maintaining coherent quantum spatial superpositions. Additionally, it has been shown that under certain conditions nanodiamonds can graphitize when optically trapped in vacuum \cite{graphitization}. Therefore, it is pertinent to explore ``dark'' trapping schemes like the magnetic trapping approach we suggest here.}

\begin{figure}[h]
    \includegraphics[width=\linewidth]{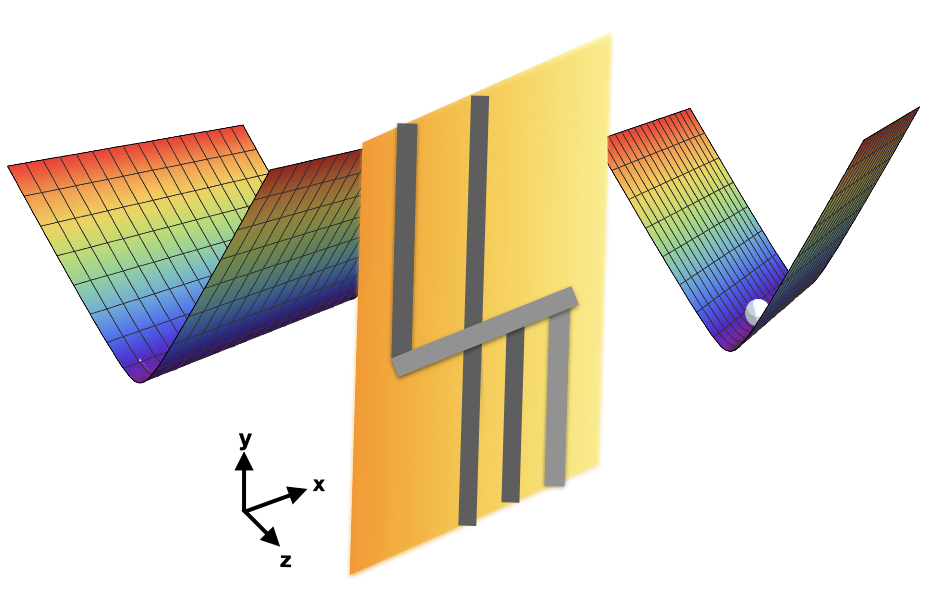}
    \caption{A schematic drawing of the two diamagnetically trapped nanodiamonds on either side of a microfabricated chip, which also functions as an electromagnetic screen, e.g. as studied in Ref. ~\cite{Schut2024}. Electric current passing through wires on the chip surface generates high-gradient magnetic fields that can be used in combination with external bias fields to diamagnetically trap nanodiamonds within distances of order $\sim 10$ $\,\si{\micro\meter}$ from the surface. 
    The rainbow-colored shapes represent the trapping potentials.
    The shape of the potential will differ slightly depending on whether wires or permanent magnets create it. In the case of wires, the setup can also be rotated horizontally; this schematic is purely for illustrative purposes.
    Although the traps are illustrated to be flat in the $\hat{x}$-direction, there can be some confinement based on the exact profile.
    e.g., a pulsed magnetic field that can be used to create spatial superpositions from the embedded spin superposition in the diamond NVC via the Stern-Gerlach effect can produce confinement in the $\hat{x}$-direction.}
    \label{fig:setup0}
\end{figure}

In this paper, we study a micro-fabricated wire-based approach inspired by techniques developed in the atom-chip community \cite{Folman2000,Fortagh2007,Amico2022,riechel}, where two nanodiamonds can be magnetically trapped in a 'long' trap with a flat direction, on either side of a microfabricated chip with an embedded (super)conducting plate. \textcolor{black}{While chip-based magnetic trapping of neutral atoms is typically accomplished  through the Zeeman effect i.e. $U_{int}$= $-\bm \mu\bf\cdot B$, where $\bm\mu$ is the magnetic moment of the atom, diamagnetic trapping of nanoparticles relies on their induced magnetic moment in the presence of an external magnetic field, yielding an interaction term $\propto \chi_V|B|^2$, where $\chi_V<0$ is the volume magnetic susceptibility of the nanoparticle.}

Using wires rather than permanent magnets to create the long trap has two useful features. First, strong magnetic field gradients can be generated close to the surface of the wires without requiring the nearby absolute magnetic field to be as large as is typical in permanent magnet-based diamagnetic traps \cite{Hsu:2016,Twamley2024}. This is advantageous for the use of thin-film superconducting (SC) magnetic shielding close to the trapping wires, as the magnetic field perpendicular to the surface of the superconductor should remain below the lower critical field to ensure the shield operates in the Meissner state.  Second, the patterned-wire-based trap allows for fast switching of magnetic fields and gradients, which is impossible with permanent magnet-based approaches.

For entanglement experiments in such a trap, a spatial superposition of each nanodiamond can be realized in the `flat' direction of a magnetic trap, with the entanglement protocol carried out with the nanoparticles confined in the transverse directions. The flat direction of the trap is such that the superposition is created parallel to the superconducting plate, as illustrated in Figure~\ref{fig:setup0}. A spatial superposition could be accomplished, for example, by applying a magnetic pulse to the spin superposition in the Nitrogen-Vacancy Centre (NVC) in the diamond.
The superconducting screen shields electromagnetic interactions between the superpositions, such that the interaction between the superpositions is via the exchange of a virtual graviton if gravity is quantum in nature.
  Our trapping approach also provides a possible route towards realizing large quantum superpositions of massive objects for other tests related to quantum coherence in macroscopic systems \cite{BassiRMP2013,Romero_Isart_2011, Bose:2017nin,vandeKamp:2020rqh,   Schut:2021svd,Schut:2024lgp}.

The remainder of this paper is organized as follows. In Sec. \ref{sec:screening}, we describe the considerations for screening of electromagnetic backgrounds in nanoparticle entanglement and superposition experiments.
In Sec. \ref{sec:wiretrapping}, we discuss the diamagnetic trapping of the diamond nanospheres using a novel wire setup, where the wires are put on a chip parallel to the superconducting plate. Here, the superconducting (conducting) plate is assumed to be made out of niobium-coated (gold-coated) silicon-nitride, and a few millimeters in length.
We also discuss the constraints on the wire magnetic field and its mirror field and their backreaction on the trapping potential in the presence of the superconducting plate. 
\textcolor{black}{Finally, in section \ref{sec:discussion}, we discuss prospects for the general creation of large quantum superpositions in traps of this type, identify sources of dephasing that would require a detailed analysis in a realistic protocol for creating superpositions in such a trapping setup, 
and discuss other possible new physics searches enabled by the proposed setup.}

\section{Screening Electromagnetic Interactions}\label{sec:screening}
To ensure gravity-induced entanglement, electromagnetic interactions between the test masses need to be screened, depending on the separation of the test masses.
The interactions between the two spherical test masses (indicated by the superscript $\text{S}-\text{S}$) in the \textit{absence} of a (super)conducting screen are the electric dipole - electric dipole (DD), Casimir-Polder (CP), magnetic dipole - magnetic dipole (MM) and gravitational (GR) interactions:
~\footnote{
We have assumed that the separation between the test masses is large compared to the radius of the spheres.
Furthermore, the test masses are assumed to be perfectly spherical and to consist solely of diamonds.
In a high vacuum environment, the polarizability of a diamond sphere can be expressed as $\alpha=R^3(\epsilon-1)/(\epsilon+2)$~\cite{kim2005static}.\label{footnote:cp}
}
~\cite{Casimir:1947kzi,griffiths2005introduction,feynman1963feynman}
\begin{align}
    V_\text{DD}^\text{S-S} &= \frac{1}{4\pi\varepsilon_0 } \left( \frac{\mathbf{d}_1\cdot\mathbf{d}_2}{r^3} - \frac{3(\mathbf{d}_1\cdot\mathbf{r})(\mathbf{d}_2\cdot\mathbf{r})}{r^5} \right) \, , \label{eq:dipole_SS}\\
    V_\text{CP}^\text{S-S} &= - \frac{23\hbar c}{4\pi} \left(\frac{\epsilon-1}{\epsilon+2} \right)^2 \frac{R^6}{r^7} \, , \label{eq:casimir_SS} \\
    V_\text{MM}^\text{S-S} &= \frac{\mu_0}{4\pi} \left( \frac{\mathbf{m}_1\cdot\mathbf{m}_2}{r^3} - \frac{3(\mathbf{m}_1\cdot\mathbf{r})(\mathbf{m}_2\cdot\mathbf{r})}{r^5} \right) \, , \label{eq:magdipole_SS} \\
    V^\text{S-S}_\text{GR} &= -\frac{G m_1 m_2}{r} \, \label{eq:grav_SS}
\end{align}
where $\epsilon=5.7$ the dielectric constant of the diamond test mass, $\varepsilon_0$ is the vacuum permittivity, $\mu_0$ is the vacuum permeability, $\hbar$ is the reduced Planck constant, and $c$ the speed of light. 
Also, $G$ is the gravitational constant and $m_1$, $m_2$ the mass of the test particles $1$ and $2$, respectively.
Furthermore, $\mathbf{r}$ is the vector connecting the center of masses of the two spheres with length $\abs{\mathbf{r}}=r$, $R$ is the radius of the test mass (which we assume to be perfect spheres), $\mathbf{d}_1$ ($\mathbf{d}_2$) is the electric dipole moment of test mass $1$ ($2$) which can consist of both a permanent electric dipole moment and an induced electric dipole moment in the case of diamond test masses, and $\mathbf{m}_1$ ($\mathbf{m}_2$) is the magnetic dipole moment of the test mass $1$ ($2$), which is induced by an external magnetic field. \newline

The electric dipole moment can be either induced or inherent to the crystal.
The permanent dipole can appear due to impurities on the surface or in the bulk.
For the diamond test masses considered here, this dipole moment has not been measured, and its scaling with mass is unknown~\footnote{
In Ref.~\cite{Afek:2021bua} the electric dipole moment of \ce{SiO_2} was measured to have no clear scaling with the volume of the material. 
However, the measurements showed at least one order of magnitude uncertainty in the dipole moment magnitude of spheres with radius $10-20\,\si{\micro\metre}$, showing that a volume scaling of the permanent dipole is an open question.
}.
In the previous paper we have taken the dipole moment to be $\abs{\textbf{d}}\sim 0.1e\,\si{\micro\metre}$ for test-mass of radius $R=500\,\si{\nano\metre}$ (corresponding to $\sim10^{-15}\,\si{\kilo\gram}$ for a spherical diamond test mass), thus assuming a volume scaling, and using the experimental data in~\cite{Afek:2021bua} as a benchmark.

Additionally, the diamond can have an induced electric dipole moment due to its dielectric properties.
The induced electric dipole of the diamond from some external electric field $\mathbf{E}$ is given by its polarizability $\alpha$ (in SI-units) via:
\footnote{
Diamond has a local polarizability $\alpha_c$, and a dipole can be induced in the atoms of the crystal due to a local electric field~\cite{jackson_classical_1999}:
\begin{equation}
    \mathbf{d}=N \alpha_\text{c} \, \mathbf{E}_{\text{loc}}.
    \label{eq:induced_63}
\end{equation}
The local polarizability of the atoms is related to the polarizability of the medium via the Claussius-Mossotti relation:
\begin{equation}
    \frac{n_v \, \alpha_\text{c}}{3 \epsilon_{0}}=\frac{\epsilon_{r}-1}{\epsilon_{r}+2} \,,
    \label{Classius-Mossotti}
\end{equation}
with $n_v$ the number density of atoms, $n_v=3N/4\pi R^3$ for spherical masses ($N$ the number of atoms, $R$ the radius) and with $\epsilon_r$ relative permittivity of the medium.
Using the Claussius-Mossotti relation in eq.~\eqref{eq:induced_63} give eq.~\eqref{eq:induced_edip}.
}
\begin{equation}\label{eq:induced_edip}
    \mathbf{d} = \alpha \mathbf{E} \, .
\end{equation}
We can consider the wires used in trapping such as discussed in section~\ref{sec:wiretrapping}, to be DC current carrying conductors such that no electric field is produced around them. 

\textcolor{black}{We consider the interaction of a spherical diamond of radius R with the external magnetic field B generated by our trap. Since diamond is a diamagnetic object, the external magnetic field can induce magnetization. If ~$R\lesssim|B|/|\nabla B|$,~we can approximate the diamond as a point dipole. Under this approximation, the magnetization, \textbf{M} and the total induced dipole moment \textbf{m} are given by}~\cite{griffiths2005introduction,jackson_classical_1999}~\footnote{\textcolor{black}{
A full treatment of the interaction energy of a finite sized diamagnetic sphere with the external magnetic field is considered in appendix \ref{appendix:bda}.}}
\textcolor{black}{\begin{equation}\label{eq:mag}
    \mathbf{M} = \frac{\chi_V}{\mu_0(1+\frac{\chi_V}{3})} \mathbf{B} 
\end{equation}
with $\chi_V$ the dimensionless volume magnetic susceptibility, which for diamond is $-2.2\times 10^{-5}$~\cite{haynes2014crc}
and,
\begin{equation}\label{eq:inducedm}
    \mathbf{m} \approx V \mathbf{M} \approx \frac{4 \pi R^3 \chi_V}{3\mu_0} \mathbf{B} \, .
\end{equation}}

\begin{figure}[t]
    \centering
    \includegraphics[width=\linewidth]{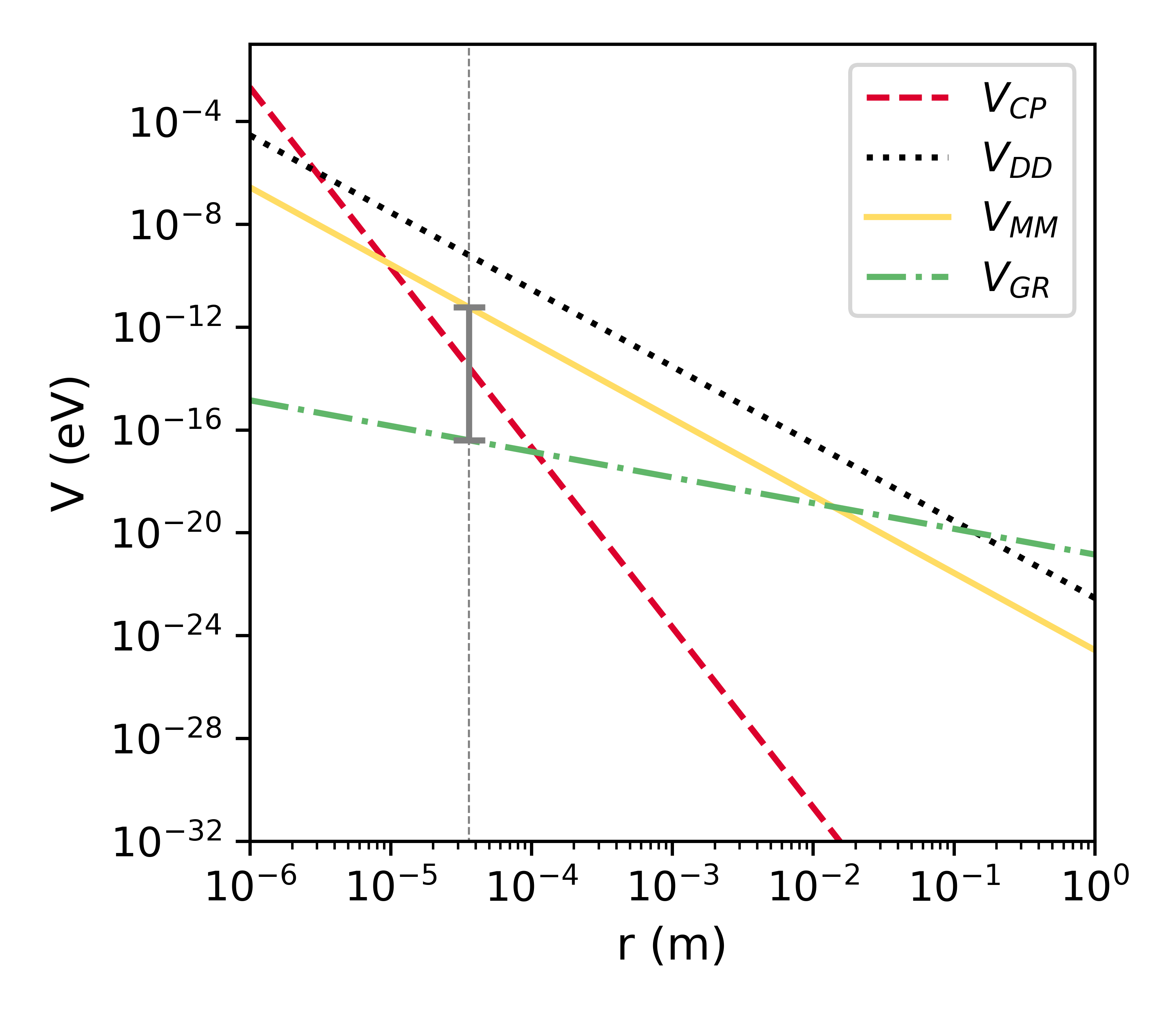}
    \caption{\small A comparison of the potentials given in eqs.~\eqref{eq:casimir_SS}-\eqref{eq:grav_SS}.
    The lines show the Casimir-Polder (CP), electric dipole-dipole (DD), magnetic dipole-dipole (MM), and gravitational (GR) interactions between two spheres with center-of-mass separation $r$.
    The test masses are considered perfect diamond nanospheres ($R=500\,\si{\nano\metre}$, $\rho = 3513 \,\si{\kilogram\per\metre\cubed}$, so $m\sim10^{-15}\,\si{\kilogram}$) with a electric dipole moment of $0.1e\,\si{ \micro\meter}$ \textcolor{black}{and a magnetic dipole moment with magnitude $\sim 10^{-18}\,\si{\joule\per\tesla}$ (the induced dipole based on a $72\,\si{\milli\tesla}$ field).
    The vertical grey line indicates a difference of approximately five orders of magnitude between the GR and MM interactions at a separation of $36\,\si{\micro\metre}$.} 
    }
    \label{fig:no-screen}
\end{figure}

Figure~\ref{fig:no-screen} shows the potentials in eqns.~\eqref{eq:dipole_SS}-\eqref{eq:grav_SS} as a function of the separation $r$.
The figure shows that the gravitational interaction dominates at large separations, while the Casimir-Polder interaction dominates at short distances.
The orientation of the dipoles is assumed to be along $\mathbf{r}$ such that the potential is maximized.
We have considered a diamond test mass with $R=500\,\si{\nano\metre}$, a dipole of $\abs{\mathbf{d}} = 0.1 e \,\si{\micro\metre}$ (from~\cite{Fragolino:2023agd}, this is the required electric dipole moment to have $\Gamma<10^{-2}\,\si{\hertz}$)  \textcolor{black}{and $\abs{\mathbf{m}} \sim 10^{-18}\,\si{\joule\per\tesla}$ (induced in a nano-diamond by a $72\,\si{\milli\tesla}$ magnetic field. This value of |B| corresponds to one of the trapping configurations considered later in the text (see Sect.~\ref{subsec:cooling}).} 
The figure shows that for the gravitational interaction to dominate the Casimir-Polder interaction tenfold; a minimal distance of $\sim150\,\si{\micro\metre}$ needs to be introduced.
This minimal distance would be $\sim50\,\si{\milli\metre}$ for the electric dipole assumed here.

In refs.~\cite{vandeKamp:2020rqh,Schut:2023eux,Schut2024} the shielding of these interactions via a conducting screen was studied.
This allowed for a reduction of the minimal separation; however, for small distances, the magnetic dipole interaction becomes dominant compared to $10\times$the gravitational interaction, as can be seen from figure~\ref{fig:no-screen} (although this depends on the expected external magnetic field).
To maximise the gravitational interaction, the separation between the test masses must be minimised.
Therefore, we consider a superconducting screen that shields the magnetic dipole and other electromagnetic interactions between the test masses.
As a consequence, there is interaction between the diamond spheres and the superconducting plate (indicated by the superscript $\text{S-P}$):
\begin{align}
    V_\text{DD}^\text{S-P} &=  - \frac{1}{4\pi\varepsilon_0}\frac{\abs{\mathbf{d}}^2}{8 z_0^3} \left[ 1+ \cos^2(\theta_e) \right]\, , \label{eq:dip_pot}\\
    V_\text{CP}^\text{S-P} &= - \frac{3\hbar c}{8\pi} \frac{\varepsilon-1}{\varepsilon+2} \frac{R^3}{z_0^4} \, , \label{eq:cp_pot} \\
    V^\text{S-P}_\text{MM} &= \frac{\mu_0 \abs{\mathbf{m}}^2}{4\pi} \frac{1}{z_0^3} [ 1+ \cos^2(\theta_m) ] \label{eq:magdip_pot} 
\end{align}
where $\theta_e$ ($\theta_m$) is the angle between the vector going from the plate to the test mass and the electric (magnetic) dipole vector, $z_0$ is the distance between the plate and the test mass ($z_0=r/2$ for an infinitesimally thin plate). The magnitude of the electric (magnetic) dipole moment is denoted $\abs{\mathbf{d}}$ ($\abs{\mathbf{m}}$).

\begin{figure}[h]
    \includegraphics[width=\linewidth]{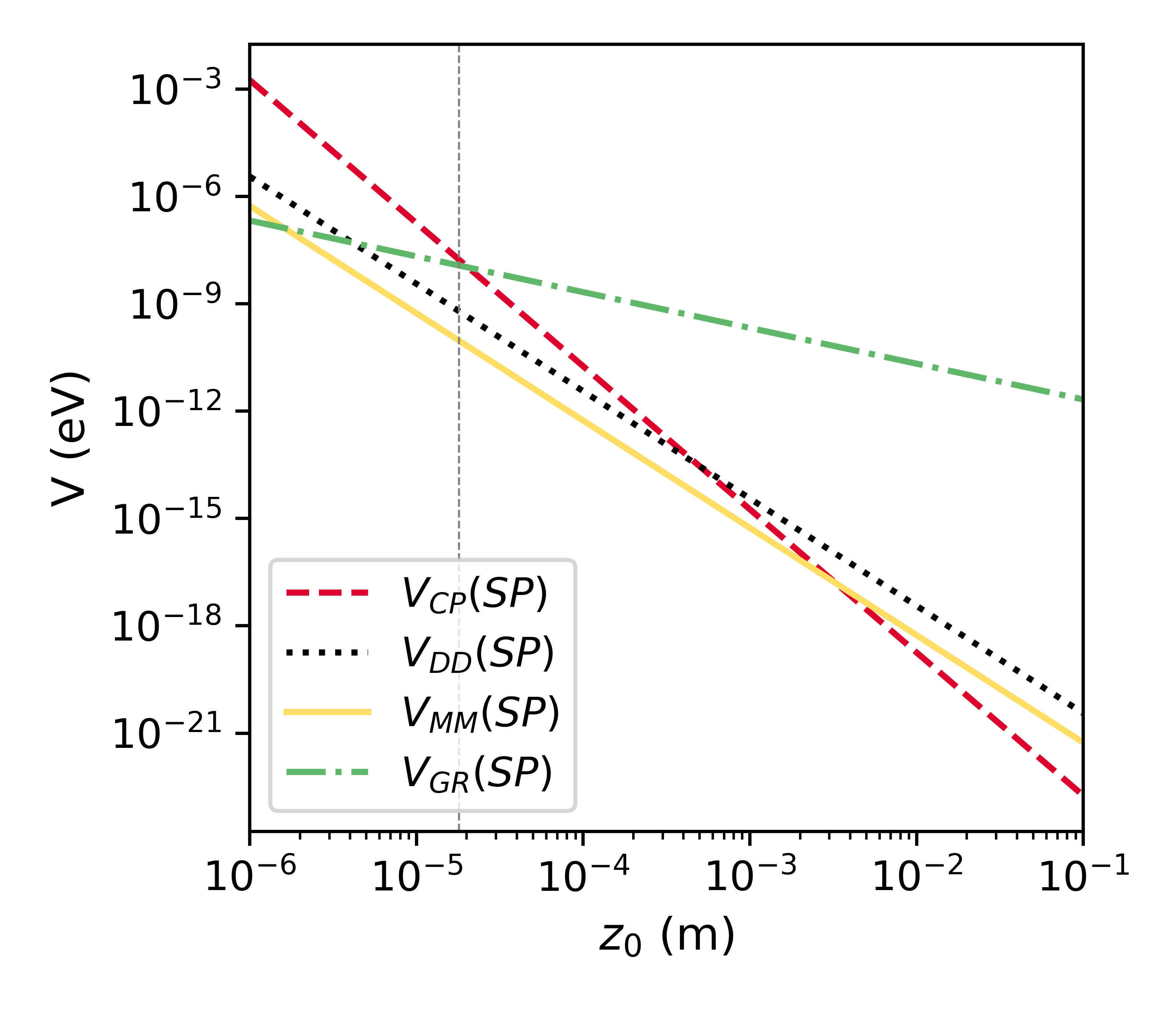}
    \caption{\small A comparison of the potentials given in eqs.~\eqref{eq:dip_pot}-\eqref{eq:magdip_pot} and eq.~\eqref{eq:grav_SS}.
    The lines show the Casimir-Polder (CP), electric dipole-dipole (DD), magnetic dipole-dipole (MM) and gravitational (GR) interactions between the sphere and the plate separated a distance $z_0$ (between center of mass). 
    The test masses are considered perfect diamond nanospheres ($R=500\,\si{\nano\metre}$, $\rho = 3513 \,\si{\kilogram\per\metre\cubed}$, so $m\sim10^{-15}\,\si{\kilogram}$) with a electric dipole moment of $0.1e\,\si{ \micro\meter}$ and a magnetic dipole moment of $\sim 10^{-18}\,\si{\joule\per\tesla}$ (the induced dipole based on a $72\,\si{\milli\tesla}$ field).
    The chip is taken to have $L=5\,\si{\milli\metre}$, and to consist of two Silicon Nitride substrates of thickness $9.5\,\si{\micro\metre}$ that squeeze a $1 \, \si{\micro\metre}$ superconducting Niobium film.
    \textcolor{black}{The dotted gray line indicates $z_0 = 18\,\si{\micro\metre}$.}
    }
    \label{fig:screen}
\end{figure}

We can also compare the sphere-plate interactions.
For gravity, we take the second mass to be that of the chip, assuming a chip to consist of two $9.5\,\si{\micro\metre}$ thick Silicon Nitride substrates with a thin $1$ micron film of Niobium (Nb).
The magnitude of the different sphere-plate interactions is shown in figure~\ref{fig:screen} as a function of the separation $z_0$.

\begin{figure}
    \centering
    \includegraphics[width=1\linewidth]{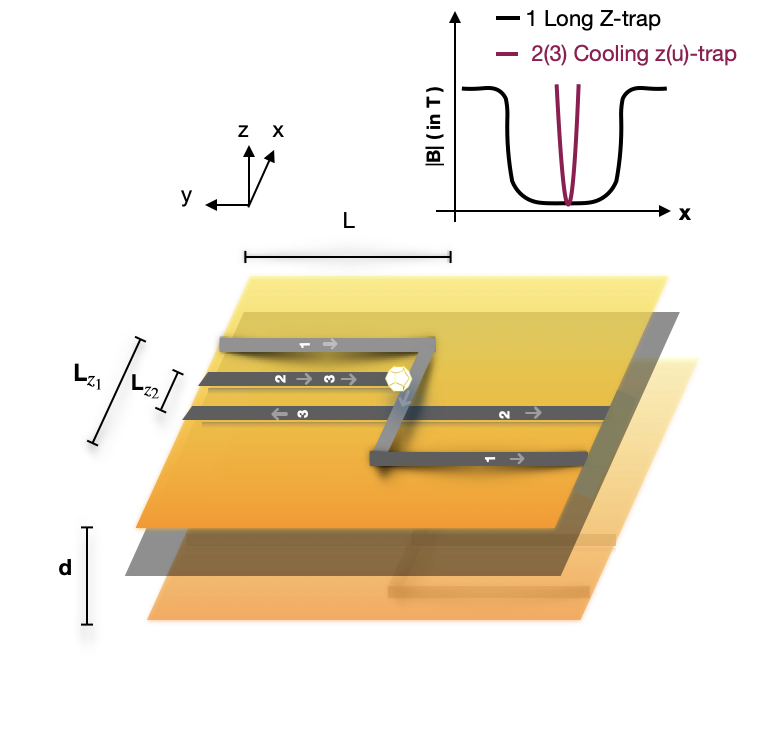}
    \caption{\small Chip-based magnetic trap with micro-fabricated wires. \textcolor{black}{The chip has three trap configurations, each made with wires of width w=5 $\,\si{\micro\meter}$}. (1) Long Z-wire with dimensions $L= 5$ mm and $L_{z_1}=1$ mm. (2) Cooling z-wire with same $L$ and  $L_{z_2}$= 30$\,\si{\micro\meter}$ and (3) Cooling u-wire with same dimensions as 2). Total \textcolor{black}{thickness} of the chip, including Nb thin film and Silicon Nitride substrate, "d"$\sim$ 20$\,\si{\micro\meter}$  
    (Origin of the coordinate system is defined w.r.t. the center of the Superconductor).  
    The particle will be cooled to its motional ground state in the stiff trap with higher frequencies. Afterwards, this trap will be switched off, and the particle will be free to move in the flat direction of the long trap, where its motional wavepacket can expand, and the Stern Gerlach pulses can be applied to create spatial superpositions. }
    \label{fig:ztrap}
\end{figure}

For the test masses to be trapped diamagnetically, the trapping potential needs to overcome the potentials given in eqs.~\eqref{eq:dip_pot}-\eqref{eq:magdip_pot}.
\textcolor{black}{The trapping potential for diamagnetic particles is given by:
\begin{equation}\label{eq:trapping_pot1}
    V_\text{T} = -\frac{1}{2}\textbf{m}\cdot\textbf{B}~
\end{equation}
Using eqs.~(\ref{eq:inducedm}) and (\ref{eq:mag}),
\begin{equation}\label{eq:trapping_pot}
    V_\text{T} = -\frac{\chi_V 4 \pi R^3 |\mathbf{B}|^2}{6\mu_0} 
\end{equation}}
Additionally, for the masses to be levitated, the trapping potential in the $\hat{y}$-direction (for the vertical chip configuration in Fig.~\ref{fig:setup0}) 
needs to overcome the earth's gravitational potential, which is given by: 
\begin{equation}\label{eq:gr1}
   V_\text{G} \approx m g y \, ,
\end{equation}
\textcolor{black}{
The nanoparticle dynamics in the trap are mainly determined by the action of diamagnetic force and the earth's gravitational pull mentioned above. 
To ensure that the particle is stably trapped, we require our trap to have a strong gradient in the y-direction (assuming a vertically oriented chip).
Electromagnetic perturbations to the trapping potential of the form of eqs.~\eqref{eq:dip_pot}-\eqref{eq:magdip_pot} are much weaker compared to the diamagnetic interaction and gravitational pull from the earth's surface (see Table~\ref{table:interaction}) and, hence, minimally affect the dynamics of the particles in the trap.}

\textcolor{black}{The particle will be trapped at the (local or global) minimum in the field.
At this minimum,
there is a force balance, i.e. (using $\bm F=-\bm \nabla V$)
\begin{align}\label{eq:grav_req}
     &m g =- \frac{\chi_V ~V}{2\mu_0} \partial_y |\mathbf{B}|^2 \,\,\text{ for the y-direction} \, \nonumber , \\
     &\text{ and} \, \nonumber , \\
     &\frac{3\abs{\mathbf{d}}^2}{32\pi \epsilon_0 z^4}[1+\cos^2(\theta_e)] + \frac{3\hbar c}{2\pi} \frac{\epsilon-1}{\epsilon+2} \frac{R^3}{z^5} \nonumber\\&\qq{}- \frac{3\mu_0}{4\pi} [1+\cos^2(\theta_m)]  \frac{\abs{\mathbf{m}}^2}{z^4} =- \frac{\chi_V V}{2\mu_0} \partial_z |\mathbf{B}|^2\nonumber\\&\text{ for the z-direction.} \nonumber\\&\qq{} 
\end{align}
which puts a minimal constraint on the strength of the magnetic field gradient.}
%
%
\textcolor{black}{Taking all interactions into account, we require a force balance as in eq.~\eqref{eq:grav_req} and also require the second derivative of the potential, which gives the frequency, to be larger than zero, so that the minimum is a stable point. 
We calculate the oscillation frequency of the particle moving under the collective effect of all the above-mentioned interactions; the frequency is given by:}
\begin{equation}
  \omega_\zeta=\left[\frac{1}{m}\frac{\partial^2V_{\mathrm{tot}}}{\partial\zeta^2}\right]^{1/2}  
\end{equation}\label{eq:freq} 
where $V_{\mathrm{tot}}$ is the potential energy due to all interactions and $\zeta=(x,y,z)$. For stable trapping, $\omega_\zeta>0$.
\begin{table}[tbh]
\centering
\renewcommand{\arraystretch}{1}
\small  
\begin{tabular}{p{4cm} p{4.5cm}}  
\toprule
Interaction & Relative strength w.r.t. diamagnetic trapping potential\\ 
\midrule
Diamagnetic&~~~~1 \\
Casimir-Polder&~~~~$10^{-7}$ \\
Electric dipole-dipole&~~~~$10^{-9}$\\
Induced Mag dipole-dipole&~~~~$10^{-10}$\\

\midrule
\bottomrule
\end{tabular}
\caption{Relative strengths of various interactions of the particle in short z trap in the vertically oriented chip (Fig.~\ref{fig:setup0}) for I= 12A and $\rm B_{\rm bias}$=200~mT. The magnetic field |B| at the equilibrium position is 72~mT, and the frequencies are given in Table.~\ref{table:freq}. The particle's position in the trap is determined by the force balance equations \eqref{eq:grav_req}. Since EM interactions \eqref{eq:dip_pot}-\eqref{eq:magdip_pot} between the particle and the plate are much weaker than the trapping potential, particle can be stably trapped under small fluctuations in these interactions. We have assumed $\theta_e,\theta_m$=0.}
\label{table:interaction}
\end{table} 

In the next section, we consider magnetic trapping based on a chip design with Z and U-shaped wires. 
We take into account the interaction between the magnetic trapping field and a superconducting plate separating the test masses and show that the above constraints are satisfied leading to stable trapping conditions.

\section{Diamagnetic trapping using Micro-fabricated Wires}\label{sec:wiretrapping}

Diamonds exhibit weak diamagnetism, which allows them to be magnetically trapped. We can fabricate magnetic traps either using permanent magnets (see, for example,  \cite{Hsu:2016,Twamley2024} ) or using current carrying wires, as is done in BEC experiments \cite{Folman2000,Fortagh2007}. In this section, we will discuss our micro-fabricated wire-based trap design suitable for trapping, cooling, spin-manipulation, splitting the center-of-mass (COM) and rapid state expansion schemes for nano to micron-sized diamonds or any other diamagnetic particles. We will use the small case "z" and "u" to represent the stiff traps and the upper case "Z" to refer the long trap.

\begin{figure} [ht]
    \centering
    \includegraphics[width=1\linewidth]{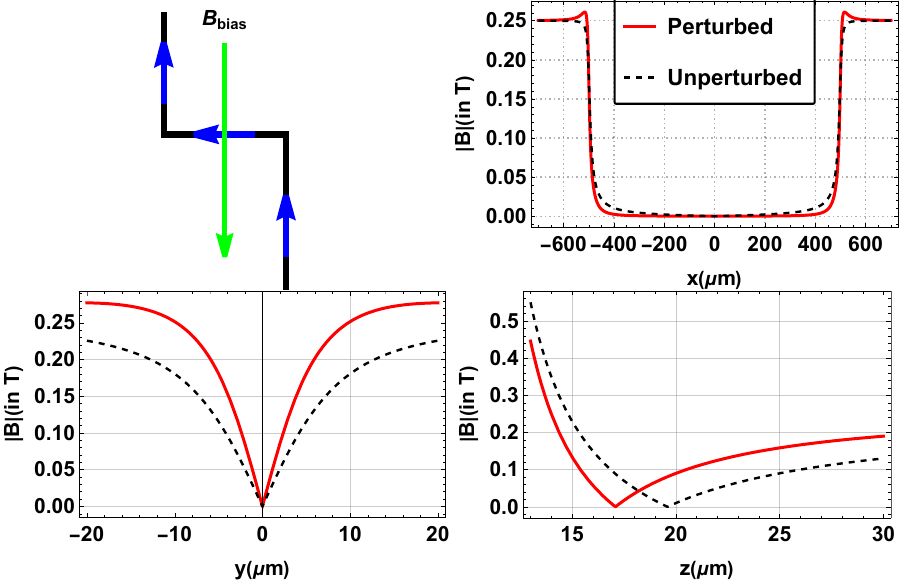}
    \caption{\small \textbf{Magnetic Field Profile of the Z-trap:} By applying a uniform magnetic bias field perpendicular to a long current-carrying wire (parallel to the SC screen), the field above the wire cancels at the center line of a long magnetic trap. It takes the form of a 2-D quadrupole field. For a chip-based magnetic trap, the magnetic trapping gradient perpendicular to the trap center line is set by the wire current $I$ and applied bias field $B_{\rm bias}$ parallel to the chip surface. The plots show the total Magnetic Field Profile (|B|) along the longitudinal(x) and transverse for the long Z trap (1mm) with current 12A and bias fields 250 mT, including the perturbation from the Meissner Image (solid red line). Height of the particle from the screen shifts from 19.6$\,\si{\micro\meter}$ to 17 $\,\si{\micro\meter}$ due to this effect. The particles are free to move along the x-axis and tightly confined along the other two directions.}\label{fig:longtrap}
\end{figure}
Wire-based setup allows in-situ manipulation of trapping parameters and provides very high magnetic field gradients for modest values of current and bias magnetic fields. In our setup, we consider "z" and "u"  shaped stiff traps for cooling the COM motion and a shallow long "Z" trap for creating non-gaussian states of our nano-diamonds (either spatial superpositions \cite{Bose:2017nin} or coherent motional wavepacket expansion~\cite{stateexpansion}). 
The wires can be fabricated onto a Silicon\textcolor{black}{/Silicon} Nitride substrate $\sim \mathcal{O}(10\,\si{\micro\meter})$.
An identical setup will be used to trap the second particle (Fig.~\ref{fig:setup0}), and to shield all EM interactions, the particles will be separated by a thin film Niobium screen operating in the Meissner state (see Fig.~\ref{fig:ztrap}).

\textcolor{black}{In order to fabricate such a thin substrate, a few micron thick SiN membrane of dimensions of a few square mm could be fabricated by depositing a low stress SiN film onto a thicker, e.g. 0.5 mm, Silicon-on-Insulator (SOI) handle wafer and then etching a pit into the thick Si handle wafer from the reverse side to form the thin SiN membrane after wire deposition \cite{Chiaverini:2003}. An insulating layer and superconducting thin film magnetic shield could also be deposited on top of the wire layer prior to releasing the membrane. Since SiN can be deposited under tension, it can form a stiff mechanically stable layer with fundamental mode frequency of order $\sim 100$ kHz \cite{Chiaverini:2003} on which to support the thin section of the chip. By affixing two such wafers front-to-front, it may be possible to realize an arrangement such as that shown in Fig. \ref{fig:setup0}, where two nanodiamonds can be trapped on opposing sides in close proximity to the thin region of a chip.}

Wires are arranged so that the currents flowing through them are parallel to the superconducting (SC) screen. We also carefully choose the current and bias field values to ensure they do not exceed the critical values for maintaining the Meissner state for thin film Niobium. For bulk Nb,
$\rm H_{c_1}(T=0)=170~mT$. The experiment can be performed in a cryogenic setup where liquid Helium is used to cool the experiment to T=4.2~K (it is possible to go to a lower temperatures). At this temperature, given the critical temperature $\rm T_c=9.25~K$ for Nb, $\rm H_c(T)=H_c(0)(1-(T/T_c)^2)=135~mT$. 
\textcolor{black}{From Fig.~\ref{fig:no-screen}, we can see that the induced magnetic dipole-dipole interaction between the nanospheres is $10^5$ stronger than their gravitational interaction at a separation of $\mathcal{O} (40\,\si{\micro\meter})$. Therefore, we need our Nb film to have a shielding efficiency of around $10^5$. This value depends on the ratio of the film thickness to the film's London penetration depth ($\lambda$). For Nb, $\lambda\approx~41\,\si{\nano\meter}$ at 4.2~K. We can achieve this shielding efficiency by a film thickness of about $\sim 750\,\si{\nano\meter} $~\cite{Fosbinder-Elkins_2022}. The exact performance heavily depends upon the experimental conditions in which the film is deposited on the substrate and the amount of impurities in the film. To account for this, we have chosen a conservative value of $1\,\si{\micro\meter}$ for the film.} Additionally, in \cite{Hudson1971}, it was shown that when the applied magnetic fields are parallel to the surface of the SC, the critical field value is much higher as compared to perpendicularly applied fields (for d/$\lambda>>1$,$\rm H_{c_1\perp}\leq H_{c_1||}\leq 1.69 H_{c_1\perp}$). We can use a similar assumption for our film. 

\begin{figure}[h]
    \centering
    \includegraphics[width=1\linewidth]{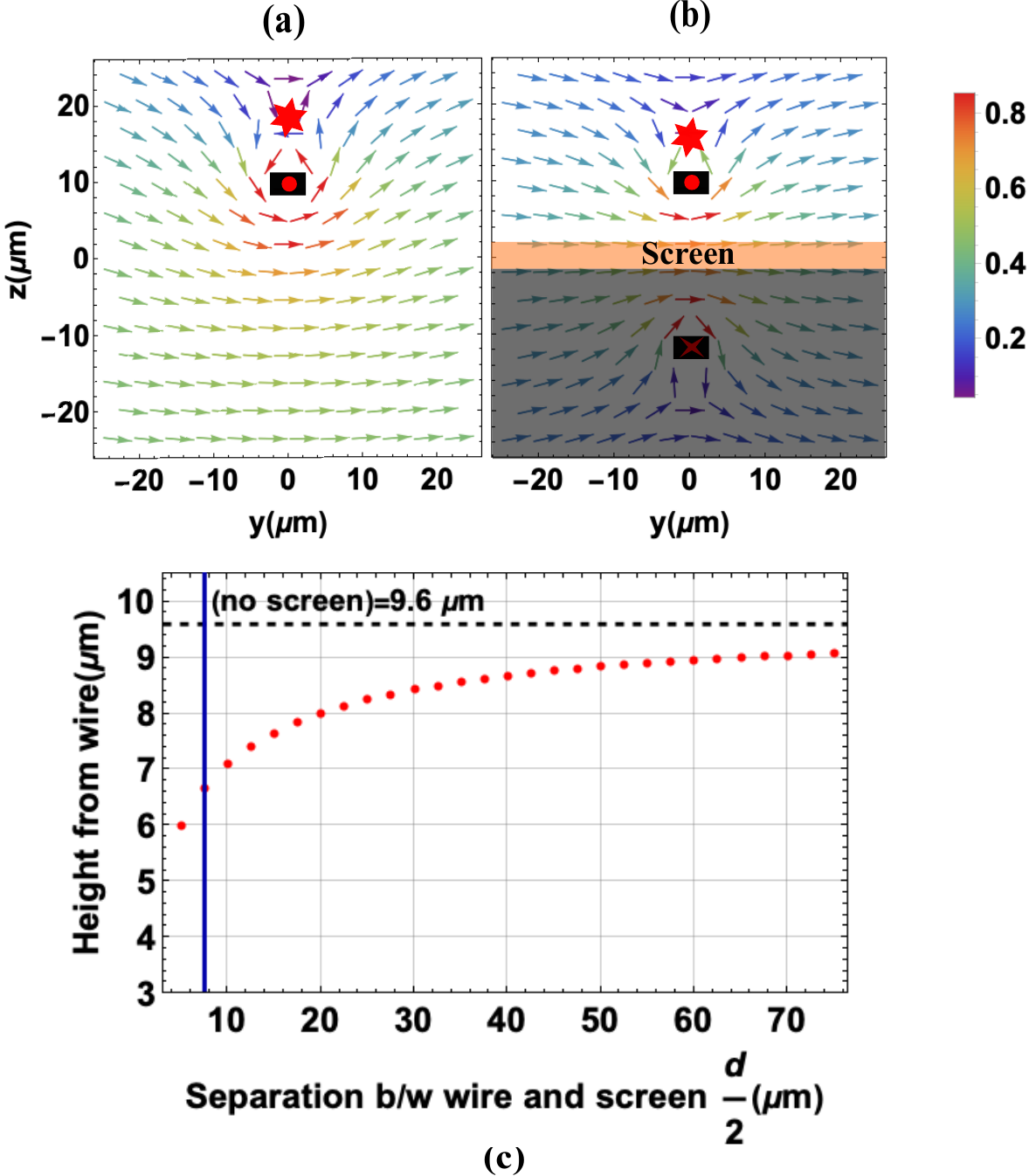}
    \caption{ When current flows through the Z-, z-and u-shaped wires, Meissner currents are generated in the SC screen to oppose the magnetic field due to the wires. This is illustrated here for the central wire using the method of images (current flows along x). \textcolor{black}{(a):} In the absence of SC screen, assuming $d/2=10\,\si{\micro\metre}$, the field minimum is created at 19.6$\,\si{\micro\metre}$ from the origin (denoted by the red star). \textcolor{black}{(b):} With the screen, field from Meissner currents perturb the field from true wires, due to which the new minimum is formed closer to the wire, at $17.1\,\si{\micro\metre}$. \textcolor{black}{(c):} Plot shows how the height of the particle from the wire decreases as the separation $d/2$ from the SC screen decreases. Closer to the SC, Meissner currents are stronger. As a result, the effective current due to the wires decreases, and the particle gets trapped closer and closer to the wire. The blue vertical line represents the thickness of the chip ($d/2> 7.5\,\si{\micro\metre}$) for which the perpendicular component of the field from the wires ($B_z$) is greater than or equal to $H_{c_1}$ for Nb at T=4.2K (135 mT). Here, I=12A, $B_{\rm bias}$=250~mT and $L_z$=1mm. }
   \label{fig:SCeffect}
\end{figure}

Due to the proximity of the wires to the SC, Meissner currents will be generated in the SC to oppose the B-field generated due to the wires which will perturb the trap (Fig.~\ref{fig:SCeffect}). 
\textcolor{black}{Additionally, since micro-fabricated wires tend to be more rectangular with height much less than their width and length, we have considered wires with finite width and zero thickness in our calculations for our traps} (magnetic field profiles are obtained using simple Biot-Savart Law calculations). Fig.~\ref{fig:ztrap} has been evaluated with all these considerations. For currents $\mathcal{O}(\rm 12)~A$ and bias fields $\mathcal{O}(\rm 0.2)$~T, we can trap the particle $\mathcal{O}(10)\,\si{\micro\meter}$ from the wires (by trapping close to the wires, we are reducing the minimum separation between the two particles, thus maximizing the entanglement phase). We can easily adjust the height $z_0$ from the screen (origin is defined w.r.t. to the center of the screen) by tuning the applied bias field (since $z_0\sim f(I/B_{\rm bias})$, see eq.~\eqref{eq:height}). The bias field can either be applied using Helmholtz coils or "D" coils on a chip (see \cite{Fosbinder-Elkins_2022}). The exact method will depend upon the particular nature of the experiment. These traps have a flat direction, x, where the particle is free to move around while being tightly confined in transverse directions.

\begin{figure}
    \centering
    \includegraphics[width=0.7\linewidth]{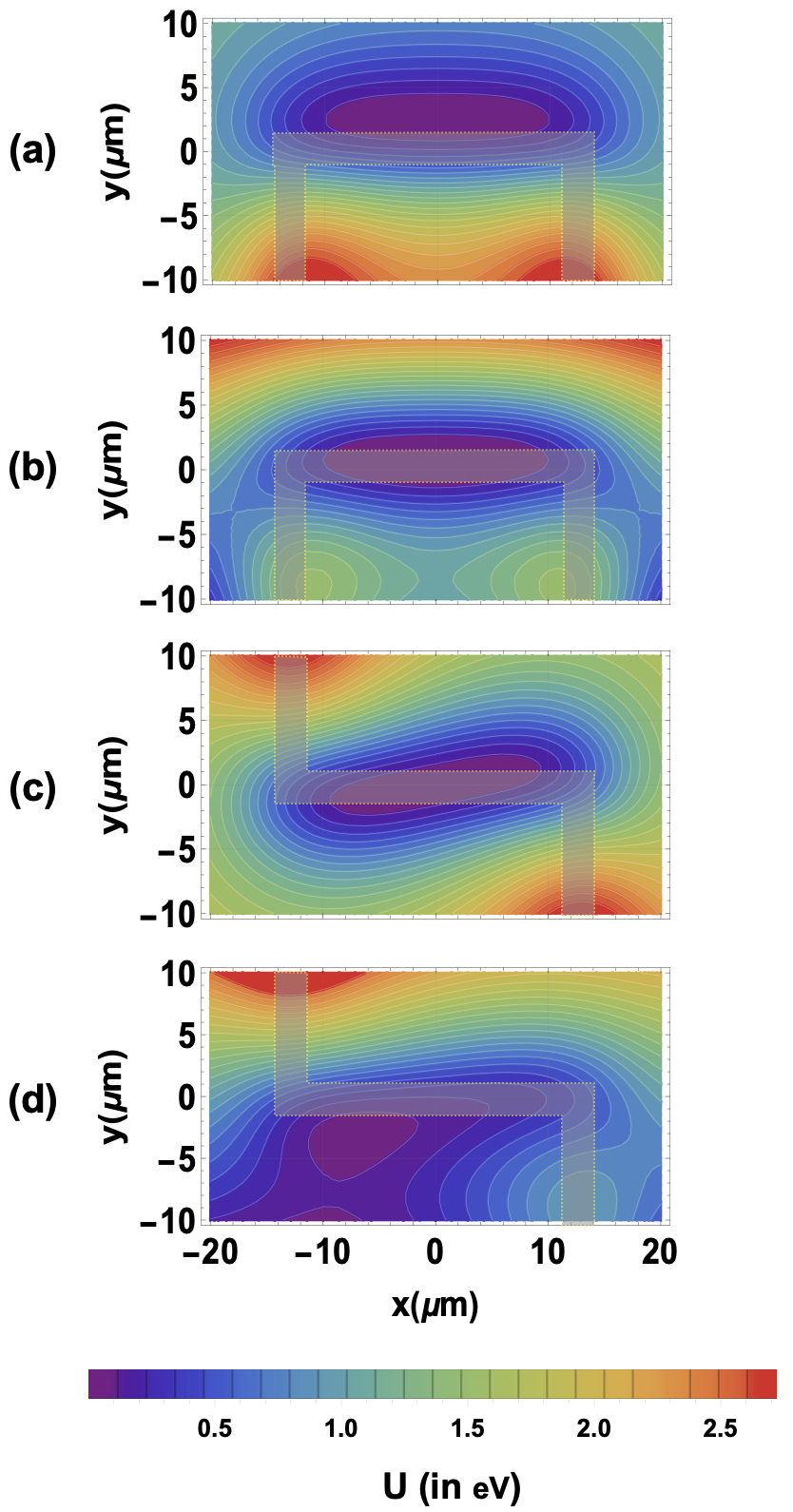}
    \caption{2D Contour plots (at respective equilibrium $z_{0}$) for potential energy (U) of the particle in short ($L_{z_1}=30\,\si{\micro\meter}$) traps in the vertical orientation of the chip (Fig.~\ref{fig:setup0}) in the lab frame. Here $B_{\rm bias}$=200~mT and I=12A. Figs. \textcolor{black}{a)} and \textcolor{black}{c)} show the energy of the particle in the diamagnetic potential for u and z traps, respectively. Figs. \textcolor{black}{b)} and \textcolor{black}{d)} show how gravity (acting along y) changes the equilibrium configuration in the u and the z settings.}
    \label{fig:contourplots}
\end{figure}

Consider a z-wire with central wire of length $L_z$ and width w of $\mathcal{O}(5)\,\si{\micro\meter}$ 
along the x-y plane parallel to the SC surface (see Fig.~\ref{fig:ztrap}). \textcolor{black}{The two particles will be positioned along the z axis on either side of the chip. The bias field and the magnetic field from the central wire will form a minimum along the direction perpendicular to the center of the wire at a height $z_0$ from the screen. 
Here, we are only accounting for the diamagnetic potential of the particle. In subsection~\ref{subsec:cooling}, we will show how gravity affects the particle's equilibrium position (see also Fig.~\ref{fig:contourplots} and Table~\ref{table:freq}).} The expressions for $z_0$ and magnetic field gradient are complicated for the short traps but can be approximated by the following expressions for the long trap.
The magnetic field magnitude at the center of the long wire at a distance d/2 from the origin (same analysis applies for the other trap at -d/2 from the origin) as a function of z is given by (assuming an infinitely long wire):
\begin{equation}\label{eq:Bapprox}
   |B(0,0,z-d/2)|= B(z)\approx\frac{\mu_0 I}{\pi w}cot^{-1}\left[\frac{2(z-d/2)}{w}\right]
\end{equation}
consistent with \cite{riechel}. The minimum height $z_0$ from the screen is
\begin{equation}\label{eq:height}
    z_0\approx d/2+\frac{w}{2}\left\{\text{cot}\left(\frac{\pi w |B_{\rm bias}|}{\mu_0 I}\right)\right\}
\end{equation}
where $B_{\rm bias}$ is the bias field applied along $-y$ to create the field minimum. The z-dependent magnetic field gradient at the center of the thick wire is given by:
 \begin{equation}\label{eq:gradient}
     \partial_z B(z)\approx -\frac{\mu_0}{2\pi}\frac{I}{(z-d/2)^2+(w/2)^2}
 \end{equation}
 which is a Lorentzian as opposed to the $1/z^2$ from the infinitessimal width wire approx.

Clearly, for $z_0-d/2\leq w$ (i.e., the separation between the wire and the nanoparticle smaller than the width of the z-wire), it becomes important to account for the width of the wire, while in another case, one can estimate $z_0$ by modeling the z-wire as an infinitessimally wide wire ($z_0$=d/2+$\mu_0$I/(2$\pi B_{\rm bias})$). Incorporating the effect of SC screening using the method of images (Fig.~\ref{fig:SCeffect}), we can see that the Meissner currents reduce the field from true wires such that the perturbed $z_0$ is smaller than the original $z_0$.


Key features of various traps will be discussed in the following subsections.

\subsection{Cooling traps}{\label{subsec:cooling}}
It is crucial to cool the COM motion of our levitated nano-diamond 
for the quantum entanglement experiments. This step could be carried out in the short "z" or "u" traps as illustrated in Fig.~\ref{fig:ztrap}. The traps are smaller in length ($\sim 30 \,\si{\micro\meter}$) to allow higher longitudinal and transverse frequencies ($\sim100\,\si{Hz}$ and $\mathcal{O}(500-800)\,\si{Hz}$) respectively as compared to the long traps. \textcolor{black}{Feedback cooling the COM motion could in principle be implemented using time-dependent magnetic field gradients applied using other wires in the chip surface or external coils, as was done in Ref. \cite{Hsu:2016}.  Alternatively, optical cooling of nanoparticles within micron range of a metallic surface has also been recently performed \cite{montoya:22,grinin:2025}.} Such high trap frequencies are advantageous for cooling closer to the quantum ground state of motion, and the longitudinal frequencies we estimate are approximately $10 \times$ higher than achieved in other permanent magnet-based long traps \cite{Hsu:2016}. In principle, we can make the cooling traps much shorter to get higher frequencies.

In the "z" configuration, we can control the $\abs{B}$ field at the center by applying an offset "Ioffe" field without changing the equilibrium position of the particle, see~\cite{Sanz:2006}. Due to gravity, the particle shifts downwards along the y direction when the chip is oriented in the vertical configuration (Fig.~\ref{fig:setup0}) and the new trap is formed closer to the upper arm of the z wire  (see  Fig.~\ref{fig:contourplots} (d)).\textcolor{black}{ The norm of the magnetic field |B| at the trap minimum is 72 mT.
The principal axes of the z -trap are tilted w.r.t. to the coordinate axes (x,y) as seen in Figs.~\ref{fig:contourplots}(c) and (d) due to the orientation of the end cap wires. Therefore, to calculate the oscillation frequencies of the particle in the trap ( table~\ref{table:freq}), we perform a coordinate transform to the principal axes where the tilt angle $\theta=-(d B_z/dx)(dB_z/dy)^{-1}$\cite{extavour}.}
 
In the case of the "u" configuration, in the absence of gravity or other external forces, the |B| field is effectively zero at the trap center. 
However, as stated above, the equilibrium position of the particle is determined by the combination of diamagnetic and gravitational forces  and the norm of the field at this point for our parameters is $\sim 62$~mT. Additionally, the potential is no longer tilted in this configuration, and particle remains centered as both end cap wires are positioned downwards and effectively counter gravity. We can trap nanoparticle in either of the two traps based on our requirements. We can also switch between the two in-situ by switching the current between top and bottom wires (2 and 3 in Fig.~\ref{fig:ztrap}).

\subsection{Long trap for creating superpositions and motional wave packet expansion}
\begin{figure}[ht]  
    \centering
    \includegraphics[width=1\linewidth]{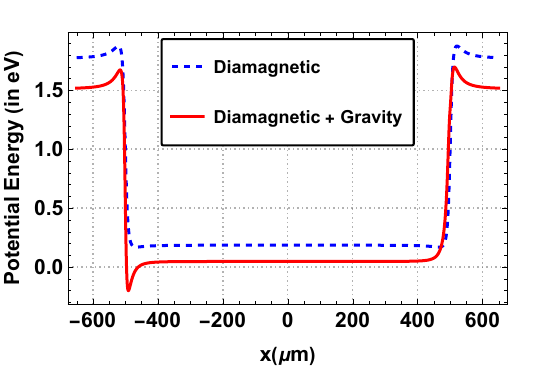}
    \caption{The minima of the long Z trap ($L_{z_1}$=1 mm) along x direction shifts from the center to one of the end wires due to gravity (for the vertically oriented chip (Fig.~\ref{fig:setup0})). It takes about $\sim$ 25$\,\si{\second}$ to roll down to the new minima. Here $B_{\rm bias}$=250~mT and $I=12$ A. }\label{fig:longtrap:GRAV}
\end{figure}

Once the particle is cooled, 
the cooling trap can be turned off to allow the motional state to expand in this long ``flat'' trap of length 1mm. 
In this trap, considering only the diamagnetic potential, the nanoparticle  oscillates with a frequency $2\pi \times 0.01$ Hz (in vertical (V) config.) along the flat x direction. Due to gravity, similar to the small z wire case discussed previously (Fig.~\ref{fig:contourplots}), the potential minima shift closer to one of the end caps (-490$\,\si{\micro\metre}$). Therefore, the particle will roll down to this new equilibrium (see Fig.~\ref{fig:longtrap:GRAV}), where it will oscillate with a frequency of 2$\pi \times 110$ Hz. The time taken by the particle to roll down to this minima is (1/4)th of its period, i.e., 25  seconds. However, in the long trap, we plan to apply Stern Gerlach pulses to create spatial superpositions and perform other operations on the particle at the center with much higher gradients at a much shorter time scale ($\sim$1 sec). Therefore, we can assume the particle doesn't roll down to this minimum. Instead, we can keep it centered while performing relevant operations on it.   From Fig.~\ref{fig:longtrap:GRAV}, we can see that the trap depth is quite high, about 1.5 eV $\sim 10^4$~K ($\gg \rm k_B T$ at T=300~K). This means the particle can be stably trapped for long durations without getting kicked out by thermal fluctuations.

\begin{table}[ht]
\centering
\renewcommand{\arraystretch}{1}
\small  
\begin{tabular}{p{1.8cm} p{2.2cm} p{3cm} p{1.2cm}}  
\toprule
\text{Trap} & \text{Orientation} & ($\mathbf{\omega_x}$, $\mathbf{\omega_y}$, $\mathbf{\omega_z}$)/2$\pi$ (Hz) & $z_{\mathrm{0}}$($\,\si{\micro\metre}$) \\
\midrule
\multicolumn{4}{c}{\text{$\rm B_{\rm bias}$= 0.2 T}} \\
\midrule
\multirow{2}{*}{\text{30$\,\si{\micro\meter}$ u-wire}} 
&~~~~~(H) & (121.5,507, 585) & 16.2 \\
&~~~~~(V) & (62, 380.7, 336) & 18.6 \\
\midrule
\multirow{2}{*}{\text{30$\,\si{\micro\meter}$ z-wire}} 
&~~~~~(H) & (191, 550, 785) & 16.4 \\
&~~~~~(V) & (159, 351, 331) & 18.3 \\
\midrule
\multicolumn{4}{c}{\text{  $\rm B_{\rm bias}$= 0.25 T}} \\
\midrule
\multirow{2}{*}{\text{1mm Z-wire}} 
&~~~~~(H) & (0.1$^*$, 553, 685) & 16.4 \\
&~~~~~(V) & (0.01$^*$,491.5, 521) & 16.7 \\
\bottomrule
\end{tabular}
\caption{Trap frequencies (along the principal axes of the trap) considering perturbations from gravity and other EM interactions for the three embedded traps. Here L=5 mm, $L_{z_1}=1\,\si{\milli\metre}$ and $L_{z_2}=30\,\si{\micro\metre}$ with Current I=12 A and d/2 = 10$\,\si{\micro\metre}$. Here $z_0$ represents the equilibrium height of the particle from the center of the SC plate. The orientations (V) and (H) correspond to whether the chip is placed vertically as per Fig.~\ref{fig:setup0} or horizontally as per Fig.~\ref{fig:ztrap}. $^{*}$Although the frequency of the particle in a long trap is given as $2\pi \times 0.1$ Hz (for H) and $2\pi \times 0.01$ Hz (for V) , we consider the case where the particle dynamics are instead determined by applied magnetic pulses.}
\label{table:freq}
\end{table}


\section{Discussion \label{sec:discussion}}
Entanglement and superposition are fundamental features of quantum mechanics, with no counterpart in classical physics. 
Using the LOCC theorem (local operations and classical communication)~\cite{Bennett:1996gf}, probing whether an interaction can entangle two massive systems prepared in quantum superpositions would prove this interaction's quantum nature. 
To experimentally test this idea for gravity \cite{Bose:2017nin} requires advances in the state of the art for creating macroscopic superpositions, scaling up from the scale of large molecules \cite{Fein2019} to micron-size particles, long spin coherence times and exquisite control over currents and magnetic fields. In this work, we proposed an integrated magnetic chip-based trap, with higher-frequency short traps (e.g. $\mathcal{O}(10 \times)$ higher in the longitudinal direction as compared to the permanent magnet trap reported in Ref. \cite{Hsu:2016}), suitable for cooling the center-of-mass motion of a nanodiamond, and a long trap with a flat direction for coherent state expansion of this motional wavepacket beyond the physical dimensions of the particle and creating spatial superpositions, for example, using a Stern Gerlach pulse sequence~\cite{Marshman:2023nkh}.  The chip incorporates an embedded superconducting shield to screen electromagnetic interactions. 
Particles can be trapped within a few microns of the chip wires, and the chip's dimensions are adjustable based on the experimental requirements. We have selected experimental parameters that are suitable for fabrication and theoretical considerations. 

\textcolor{black}{An entanglement experiment in such a setup would still be demanding technologically. We envisage typically the following stages: 1) trapping and cooling, 2) creating a macroscopic superposition and recombining the trajectories, and 3) investigating gravity-mediated entanglement with two adjacent nanodiamond superpositions. At each stage of the experiment, it is essential to account for the relevant noise sources and design appropriate mitigation strategies. In the first stage of the experiment i.e. trapping and cooling using our chip-based setup, vibrational and technical noise can couple from the mechanical tethering of the chip and affect the dynamics of the particle. It is possible to substantially mitigate this vibrational noise using inverted pendula and geometric anti-spring filter-based vibrational isolation systems (e.g. see \cite{grinin:2025}). We do not provide any detailed estimates here since the requirements on the initial conditions depend upon the final strategies for implementing steps 2) and 3) which are currently under development. Furthermore, we should also take into account of the current fluctuations, which will be among the main noise sources for the trap as well as a point of concern from the dephasing the interferometer \cite{Moorthy:2025fnu}.}
Having a trapping and screening approach in principle permits having the two nanodiamonds to be closer together resulting in a larger entanglement rate and thus requiring a smaller superposition size \cite{Schut:2023eux,Schut2024}.
However, dephasing due to interactions with the superconducting screen via the Casimir-Polder, Van der Waals, electric and magnetic dipole, current fluctuations and thermal fluctuations have to be considered. Such an analysis of the experimental parameters and requirements on control and fluctuations in the state preparation was performed in Ref.~\cite{Schut:2023eux} for an electromagnetically screened configuration, and a similar analysis for the trapped and screened protocol for a nanodiamond entanglement experiment in present setup is left for future work. \textcolor{black}{Analysis for a Stern-Gerlach pulsed sequence is also left for future study, including designing a wire-based Stern-Gerlach magnetic pulse sequence for creating spatial superposition of the nanodiamond, evaluating the trajectories of these particles, and analyzing fluctuations. The mathematical framework for the noise analysis in similar setups is well-known \cite{Moorthy:2025fnu}. However, tuning that to our requirements for a particular nanoparticle entanglement experiment will need further analysis, which we leave for a future investigation.}

While we have mainly discussed the features of the chip-based approach in the context of 
gravity-mediated entanglement experiments, the idea can be extended to other entanglement-based experiments (e.g. magnetic dipole mediated entanglement \cite{Marshman:2023nkh}), tests of quantum coherence \cite{BassiRMP2013}, search for fifth forces and axions~\cite{Barker:2022mdz}, modified theories of gravity~\cite{Elahi:2023ozf,Chakraborty:2023kel}  and in general, to the experiments for creating non-gaussian motional states of a particle. 
The wire-based chip design also provides a possible route to implement Stern-Gerlach pulses and exponential expansion of a nanodiamond wavepacket by creating an inverted harmonic potential~\cite{Romero_Isart_2017} (see also \cite{zhou2024spindependentforceinvertedharmonic}-\cite{braccini2024exponentialexpansionmassiveschrodinger}) using tailored magnetic fields and magnetic field gradients produced by appropriately patterned circuits.

Finally, the present setup can be used to investigate new short-range forces predicted to occur in some theories of physics beyond the standard model \cite{Geraci:2010,moore2021searching, Elahi:2023ozf}. There is a vast 17-order-magnitude disparity between the apparent energy scale of quantum gravity and that of the other Standard Model (electro-weak) forces.  However, several recent theories have suggested that important clues related to this “hierarchy problem” can be obtained in low-energy experiments by measuring how gravity behaves at sub-millimeter distances \cite{GiudiceDimopoulos}. Another motivation for novel short-range forces comes from the dark energy length scale \cite{Montero2023, kapner2007}.  

\section{Appendix}

\subsection{Beyond point dipole approximation}\label{appendix:bda}
In this section, we will perform a more rigorous analysis to understand the interaction of a finite diamagnetic sphere with the non-uniform external field B.

If the scale at which the norm of the magnetic field changes is larger than the physical dimension of the particle, dipole approximation holds. In our case, we quantify this by constructing a magnetic length scale $L_B$,
\begin{equation}\label{eq:scale}
    L_B=\frac{|B|}{|\nabla B|}>>R
\end{equation}
where |B| is the norm of the magnetic field. For the cooling z wire,~$L_B=1.8~\mu m>> R$ ($0.5 \mu m$), so the dipole approximation should hold. We again note that the trap location is mainly determined by the force balance between diamagnetic energy and gravity and at this point $|\nabla B|\ne0$.

 In the next section, we will derive the potential energy and the frequencies of a finite sized diamond nanosphere in an external B field and study the case where finite size effects become important.
\subsubsection{Evaluating Potential Energy}
General expression for the interaction energy of a localized magnetization with the field that induces the magnetization is given by \cite{jackson_classical_1999} :
\begin{equation}
    U=-\frac{1}{2}\int_{V}dV~\bm {M \cdot B}\approx-\frac{\chi_V}{2\mu_0}\int_V dV~ |\bm{B}|^2
\end{equation}
where M is the magnetization (magnetic moment per unit volume), $\bm{M}\approx \frac{\chi_V}{\mu_0} \bm{B}$ and $\bm{B}$ is the static external field from the wire configuration. The integration is over the volume of the sphere. Assuming the B field varies slowly over the scale of the particle, we can therefore expand the magnetic field about the trap center, keeping only upto the second order terms:
\begin{align}
    B_i(q,q_e)\approx & B_{i}(q_e)+\sum_j\frac{\partial B_i}{\partial q_j}(q_j-q_{e,j})\\\nonumber
    &+\frac{1}{2}\sum_{j,k}\frac{\partial^2 B_i}{\partial q_j\partial q_k}(q_j-q_{e,j})(q_k-q_{e,k})
    \end{align}

where $q_e=\{x_e,y_e,z_e\}$ are the equilibrium coordinates calculated from the point dipole force balance (\ref{table:freq}), $B_i$ represents the ith component of the magnetic field, and i,j,k $\in$ {1,2,3} run over the spatial directions $\{x,y,z\}$.

Assume the particle is slightly displaced to $(x^\prime,y^\prime,z^\prime)$ from the equilibrium position $(x_e,y_e,z_e)$ such that $\delta x=(x_e-x^\prime)$, $\delta y=(y_e-y^\prime)$ and~$ \delta z=(z_e-z^\prime)$ represent small displacements from the equilibrium point.  To obtain the energy U, we perform a coordinate transform from cartesian to spherical polar coordinate system centered around $(x^\prime,y^\prime,z^\prime)$ such that 
\begin{align}
    x=&r~sin(\theta)cos(\phi)+x^\prime\\\nonumber
    y=&r~sin(\theta)sin(\phi)+y^\prime\\\nonumber
    z=&r~cos(\theta)+z^\prime
\end{align}
where $r\in [0,R],\theta\in [0,\pi], \phi \in [0,2\pi]$.~Using this, we can evaluate the integral over the volume of the sphere of radius R. Keeping terms upto $\mathcal{O}$($\delta x^2,\delta y^2,\delta z^2)$,
\begin{widetext}
\begin{multline}\label{eq:finalpotential}
U = -\frac{\chi_V V}{2\mu_0} \bigg[ \;
\bigg\{\sum_iB_{0i}^2 -2\sum_{i,j}  B_{0i}B_{i,j}\delta j+2\sum_{\substack{i,j,k\\ j\neq k}}B_{i,j}B_{i,k}\delta j\delta k+  \sum_{i,j}\bigg(B_{0i} B_{i,jj}+B_{i,j}^2\bigg)\delta j^2\bigg\}\\[0.5em]
+ \frac{\textbf{R}^2}{5}\bigg\{ \sum_{i,j}\bigg(B_{0i} B_{i,jj}+B_{i,j}^2\bigg)-\bigg(3\sum_{i,j}  B_{i,jj}B_{i,j}+\sum_{\substack{i,j,k\\ j\neq k}}B_{i,j}B_{i,kk}\bigg)\delta j+ \bigg(\frac{3}{2}\sum_{i,j} B_{i,jj}^2+\frac{1}{2}\sum_{\substack{i,j,k\\ j\neq k}} B_{i,jj}\,B_{i,kk}
\bigg)\delta j^2 \bigg\}\\[0.5em]
+ \frac{\textbf{R}^4}{140} \bigg( 3\sum_{i,j}  B_{i,jj}^2+\sum_{\substack{i,j,k\\ j\neq k}} B_{i,jj}\,B_{i,kk}\bigg)
\bigg]
\end{multline}
\end{widetext}

  \begin{figure*} [ht]
    \centering
    \includegraphics[width=\textwidth]{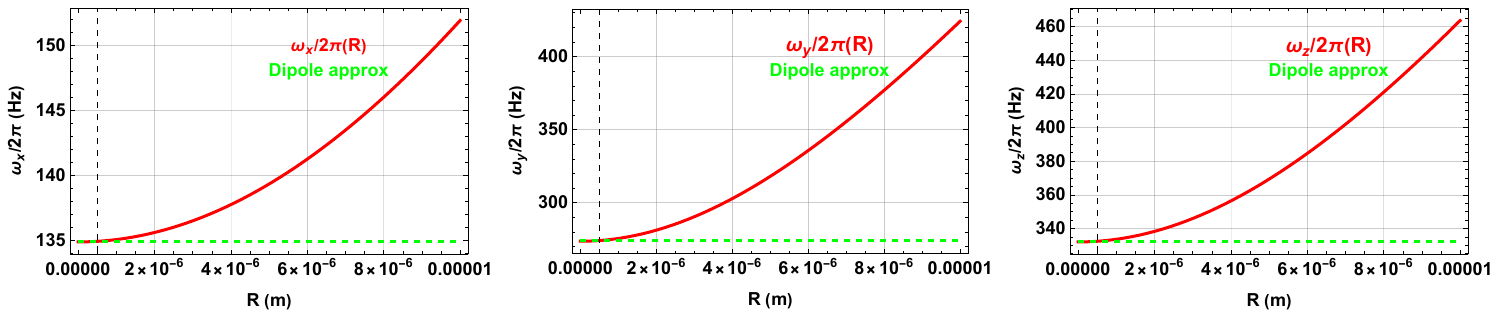}
    \caption{Oscillation frequencies (in the lab frame) of the particle in the short z trap as a function of radius R. The green dashed line shows the frequency obtained using dipole approximation and the red solid line shows how the frequency scales with the size of the particle, as seen in eq.~\ref{eq:truefreq}. The black dashed line shows R=500~nm corresponding to the system discussed in the paper. \label{fig:oscx}}
\end{figure*}

\textit{Definitions:}
The potential \( U \) describes the interaction energy of a diamagnetic object in an inhomogeneous magnetic field, expanded to second order in displacements \( \delta x, \delta y, \delta z \) from the equilibrium point \( (x_e, y_e, z_e) \). The following definitions apply:

\begin{itemize}
    \item The indices i,j,k run over the spatial distances {x,y,z}.  \( \delta x, \delta y, \delta z \) represent small displacements from the equilibrium position along the \( x \), \( y \), and \( z \) axes, respectively.
    
    \item $ B_{0i}$: Components of the magnetic field \( B_x, B_y, B_z \) evaluated at the equilibrium point \( (x_e, y_e, z_e) \).    
    \item \( B_{i,j} \): First partial derivative of the \( i \)-th magnetic field component with respect to the \( j \)-th spatial coordinate, evaluated at equilibrium. For example, 
    \( B_{z,x} = \left. \partial B_z/\partial x \right|_{(x_e, y_e, z_e)} \).    
    \item \( B_{i,jk} \): Second partial derivative of the \( i \)-th field component with respect to \( j \) and \( k \), also evaluated at equilibrium. For example, 
    \( B_{x,yy} = \left. \partial^2 B_x/\partial y^2\right|_{(x_e, y_e, z_e)} \).    
\end{itemize}

The expansion includes all terms up to second order in displacements. Linear terms represent force contributions, while second-order terms determine the curvature of the potential and relate to effective trapping frequencies.

\subsubsection{Trap frequencies}
The frequency is defined by Eq.~\eqref{eq:freq}. Using the expression for the potential energy above, we can see the explicit dependence of R in the frequencies:
\begin{widetext}
\begin{equation}\label{eq:truefreq}
    \omega_{ln}=\Bigg[\frac{-\chi_V V}{2\mu_0 m}\bigg\{4\sum_{\substack{i}}B_{i,n}B_{i,l}\big(1-\delta_{n,l}\big)+  2\sum_{i}\bigg(B_{0i} B_{i,nn}+B_{i,n}^2\bigg)\delta_{n,l}+\frac{\textbf{R}^2}{5}\bigg(3\sum_{i} B_{i,nn}^2+\sum_{\substack{i,k\\ k\neq n}} B_{i,nn}\,B_{i,kk}
\bigg)\delta_{n,l} \bigg\}\Bigg]^{1/2}
\end{equation}
\end{widetext}
Here (n,l) run over the physical dimensions (x,y,z). $\delta_{n,l}$ is the Kronecker delta (not to be confused with the notation for displacements $\delta j$)~and is non-zero only when n=l. \footnote{\small {Note that Eq.~\ref{eq:truefreq} gives us the three dimensional matrix elements for the frequency of the nanoparticle in a lab frame. However, for brevity, throughout the paper, we recast the frequency modes as $\omega_x,~\omega_y, \omega_z$. The correct interpretation for a rigid body should be $\omega_i \equiv \omega_{ii}$, where $i=x,~y,~z$. }}


In Fig.~\ref{fig:oscx}, we consider the particle in the short z trap and compare the frequencies obtained using dipole approximation and the method above. For our case, there is agreement between the two methods and hence, we can neglect the finite size effects. However, as the size of the particle gets comparable or larger than the characteristic length scale of the field $L_B$, dipole approximation starts to fail and it becomes important to consider the full interaction energy.

\section{Acknowledgements}
S.G.E would like to thank Xing Fan and Alex Hipp for helpful discussions on thin film superconductors.  AG is supported in part by NSF grants PHY-2110524 and PHY-2111544, the Heising-Simons Foundation, the W.M. Keck Foundation, , the John Templeton Foundation, DARPA, and ONR Grant N00014-18-1-2370. SB
thanks EPSRC grants EP/R029075/1, EP/X009467/1,
and ST/W006227/1.
S.B., A.G. and A.M.'s research is funded by the Gordon and Betty Moore Foundation through Grant GBMF12328, DOI 10.37807/GBMF12328, and the Alfred P. Sloan Foundation under Grant No. G-2023-21130

\bibliography{bibliography.bib} 

\begin{thebibliography}{10}

\bibitem{dyson_is_2013}
F.~Dyson, ``{IS} {A} {GRAVITON} {DETECTABLE}?,'' {\em International Journal of Modern Physics A}, vol.~28, p.~1330041, Oct. 2013.

\bibitem{Bose:2017nin}
S.~Bose, A.~Mazumdar, G.~W. Morley, H.~Ulbricht, M.~Toro\v{s}, M.~Paternostro, A.~A. Geraci, P.~F. Barker, M.~S. Kim, and G.~Milburn, ``{Spin Entanglement Witness for Quantum Gravity},'' {\em Phys. Rev. Lett.}, vol.~119, no.~24, p.~240401, 2017.

\bibitem{ICTS}
\url{https://www.youtube.com/watch?v=0Fv-0k13s_k}, 2016.
\newblock Accessed 1/11/22.

\bibitem{Marletto:2017kzi}
C.~Marletto and V.~Vedral, ``{Gravitationally-induced entanglement between two massive particles is sufficient evidence of quantum effects in gravity},'' {\em Phys. Rev. Lett.}, vol.~119, no.~24, p.~240402, 2017.

\bibitem{Horodecki:2009zz}
R.~Horodecki, P.~Horodecki, M.~Horodecki, and K.~Horodecki, ``{Quantum entanglement},'' {\em Rev. Mod. Phys.}, vol.~81, pp.~865--942, 2009.

\bibitem{Marshman:2019sne}
R.~J. Marshman, A.~Mazumdar, and S.~Bose, ``{Locality and entanglement in table-top testing of the quantum nature of linearized gravity},'' {\em Phys. Rev. A}, vol.~101, no.~5, p.~052110, 2020.

\bibitem{Bose:2022uxe}
S.~Bose, A.~Mazumdar, M.~Schut, and M.~Toro\v{s}, ``{Mechanism for the quantum natured gravitons to entangle masses},'' {\em Phys. Rev. D}, vol.~105, no.~10, p.~106028, 2022.

\bibitem{Vinckers:2023grv}
U.~K. Beckering~Vinckers, A.~de~la Cruz-Dombriz, and A.~Mazumdar, ``Quantum entanglement of masses with nonlocal gravitational interaction,'' {\em Phys. Rev. D}, vol.~107, p.~124036, Jun 2023.

\bibitem{Gupta}
S.~N. Gupta, ``Quantization of einstein's gravitational field: Linear approximation,'' {\em Proceedings of the Physical Society. Section A}, vol.~65, p.~161, mar 1952.

\bibitem{Donoghue:1994dn}
J.~F. Donoghue, ``{General relativity as an effective field theory: The leading quantum corrections},'' {\em Phys. Rev. D}, vol.~50, pp.~3874--3888, 1994.

\bibitem{Carney_2019}
D.~Carney, P.~C.~E. Stamp, and J.~M. Taylor, ``Tabletop experiments for quantum gravity: a user's manual,'' {\em Class. Quant. Grav.}, vol.~36, p.~034001, 2019.

\bibitem{Carney:2021vvt}
D.~Carney, ``{Newton, entanglement, and the graviton},'' {\em Phys. Rev. D}, vol.~105, no.~2, p.~024029, 2022.

\bibitem{Danielson:2021egj}
D.~L. Danielson, G.~Satishchandran, and R.~M. Wald, ``{Gravitationally mediated entanglement: Newtonian field versus gravitons},'' {\em Phys. Rev. D}, vol.~105, p.~086001, 2022.

\bibitem{Christodoulou:2022mkf}
M.~Christodoulou, A.~Di~Biagio, M.~Aspelmeyer, C.~Brukner, C.~Rovelli, and R.~Howl, ``{Locally Mediated Entanglement in Linearized Quantum Gravity},'' {\em Phys. Rev. Lett.}, vol.~130, no.~10, p.~100202, 2023.

\bibitem{Elahi:2023ozf}
S.~G. Elahi and A.~Mazumdar, ``Probing massless and massive gravitons via entanglement in a warped extra dimension,'' {\em Phys. Rev. D}, vol.~108, p.~035018, Aug 2023.

\bibitem{Biswas:2022qto}
D.~Biswas, S.~Bose, A.~Mazumdar, and M.~Toro\ifmmode~\check{s}\else \v{s}\fi{}, ``Gravitational optomechanics: Photon-matter entanglement via graviton exchange,'' {\em Phys. Rev. D}, vol.~108, p.~064023, Sep 2023.

\bibitem{vandeKamp:2020rqh}
T.~W. van~de Kamp, R.~J. Marshman, S.~Bose, and A.~Mazumdar, ``{Quantum Gravity Witness via Entanglement of Masses: Casimir Screening},'' {\em Phys. Rev. A}, vol.~102, no.~6, p.~062807, 2020.

\bibitem{Schut2024}
M.~Schut, A.~Geraci, S.~Bose, and A.~Mazumdar, ``Micrometer-size spatial superpositions for the qgem protocol via screening and trapping,'' {\em Phys. Rev. Res.}, vol.~6, p.~013199, Feb 2024.

\bibitem{Schut:2023eux}
M.~Schut, A.~Grinin, A.~Dana, S.~Bose, A.~Geraci, and A.~Mazumdar, ``{Relaxation of experimental parameters in a quantum-gravity-induced entanglement of masses protocol using electromagnetic screening},'' {\em Phys. Rev. Res.}, vol.~5, no.~4, p.~043170, 2023.

\bibitem{Casimir:1947kzi}
H.~B.~G. Casimir and D.~Polder, ``{The Influence of retardation on the London-van der Waals forces},'' {\em Phys. Rev.}, vol.~73, pp.~360--372, 1948.

\bibitem{Casimir:1948dh}
H.~B.~G. Casimir, ``{On the Attraction Between Two Perfectly Conducting Plates},'' {\em Indag. Math.}, vol.~10, pp.~261--263, 1948.

\bibitem{Speake}
C.~C. Speake and C.~Trenkel, ``Forces between conducting surfaces due to spatial variations of surface potential,'' {\em Phys. Rev. Lett.}, vol.~90, p.~160403, Apr 2003.

\bibitem{Kim:2009mr}
W.-J. Kim, A.~O. Sushkov, D.~A.~R. Dalvit, and S.~K. Lamoreaux, ``{Surface Contact Potential Patches and Casimir Force Measurements},'' {\em Phys. Rev. A}, vol.~81, p.~022505, 2010.

\bibitem{Behunin_2012}
R.~O. Behunin, Y.~Zeng, D.~A.~R. Dalvit, and S.~Reynaud, ``Electrostatic patch effects in casimir-force experiments performed in the sphere-plane geometry,'' {\em Phys. Rev. A}, vol.~86, Nov. 2012.

\bibitem{Schut:2023tce}
M.~Schut, H.~Bosma, M.~Wu, M.~Toro\v{s}, S.~Bose, and A.~Mazumdar, ``{Dephasing due to electromagnetic interactions in spatial qubits},'' {\em Phys. Rev. A}, vol.~110, no.~2, p.~022412, 2024.

\bibitem{Fragolino:2023agd}
P.~Fragolino, M.~Schut, M.~Toro\v{s}, S.~Bose, and A.~Mazumdar, ``{Decoherence of a matter-wave interferometer due to dipole-dipole interactions},'' {\em Phys. Rev. A}, vol.~109, no.~3, p.~033301, 2024.

\bibitem{delic2020cooling}
U.~Deli{\'c}, M.~Reisenbauer, K.~Dare, D.~Grass, V.~Vuleti{\'c}, N.~Kiesel, and M.~Aspelmeyer, ``Cooling of a levitated nanoparticle to the motional quantum ground state,'' {\em Science}, vol.~367, no.~6480, pp.~892--895, 2020.

\bibitem{tebbenjohanns2021quantum}
F.~Tebbenjohanns, M.~L. Mattana, M.~Rossi, M.~Frimmer, and L.~Novotny, ``Quantum control of a nanoparticle optically levitated in cryogenic free space,'' {\em Nature}, vol.~595, no.~7867, pp.~378--382, 2021.

\bibitem{graphitization}
A.~Rahman, A.~Frangeskou, M.~S. Kim, and et~al., ``Burning and graphitization of optically levitated nanodiamonds in vacuum.,'' {\em Scientific Reports}, no.~6, 2016.

\bibitem{Folman2000}
R.~Folman, P.~Kr\"uger, D.~Cassettari, B.~Hessmo, T.~Maier, and J.~Schmiedmayer, ``Controlling cold atoms using nanofabricated surfaces: Atom chips,'' {\em Phys. Rev. Lett.}, vol.~84, pp.~4749--4752, May 2000.

\bibitem{Fortagh2007}
J.~Fort\'agh and C.~Zimmermann, ``Magnetic microtraps for ultracold atoms,'' {\em Rev. Mod. Phys.}, vol.~79, pp.~235--289, Feb 2007.

\bibitem{Amico2022}
L.~Amico, D.~Anderson, M.~Boshier, J.-P. Brantut, L.-C. Kwek, A.~Minguzzi, and W.~von Klitzing, ``Colloquium: Atomtronic circuits: From many-body physics to quantum technologies,'' {\em Rev. Mod. Phys.}, vol.~94, p.~041001, Nov 2022.

\bibitem{riechel}
J.~Reichel, ``Microchip traps and bose–einstein condensation,'' {\em Applied Physics B}, vol.~74, April 2002.

\bibitem{Hsu:2016}
J.-F. Hsu, P.~Ji, C.~W. Lewandowski, and B.~D’Urso, ``Cooling the motion of diamond nanocrystals in a magneto-gravitational trap in high vacuum,'' {\em Scientific reports}, vol.~6, no.~1, p.~30125, 2016.

\bibitem{Twamley2024}
S.~Tian, K.~Jadeja, D.~Kim, A.~Hodges, G.~C. Hermosa, C.~Cusicanqui, R.~Lecamwasam, J.~E. Downes, and J.~Twamley, ``{Feedback cooling of an insulating high-Q diamagnetically levitated plate},'' {\em Applied Physics Letters}, vol.~124, p.~124002, 03 2024.

\bibitem{BassiRMP2013}
A.~Bassi, K.~Lochan, S.~Satin, T.~P. Singh, and H.~Ulbricht, ``Models of wave-function collapse, underlying theories, and experimental tests,'' {\em Rev. Mod. Phys.}, vol.~85, pp.~471--527, Apr 2013.

\bibitem{Romero_Isart_2011}
O.~Romero-Isart, ``Quantum superposition of massive objects and collapse models,'' {\em Phys. Rev. A}, vol.~84, nov 2011.

\bibitem{Schut:2021svd}
M.~Schut, J.~Tilly, R.~J. Marshman, S.~Bose, and A.~Mazumdar, ``{Improving resilience of quantum-gravity-induced entanglement of masses to decoherence using three superpositions},'' {\em Phys. Rev. A}, vol.~105, no.~3, p.~032411, 2022.

\bibitem{Schut:2024lgp}
M.~Schut, P.~Andriolo, M.~Toro\v{s}, S.~Bose, and A.~Mazumdar, ``{Decoherence rate expression due to air molecule scattering in spatial qubits},'' {\em arXiv}, vol.~[quant-ph:2410.20910], 2024.

\bibitem{kim2005static}
H.-Y. Kim, J.~O. Sofo, D.~Velegol, M.~W. Cole, and G.~Mukhopadhyay, ``Static polarizabilities of dielectric nanoclusters,'' {\em Phys. Rev. A}, vol.~72, no.~5, p.~053201, 2005.

\bibitem{griffiths2005introduction}
D.~J. Griffiths, ``Introduction to electrodynamics,'' 2005.

\bibitem{feynman1963feynman}
R.~Feynman, R.~Leighton, and M.~Sands, ``Feynman lectures on physics. vol. 2: Mainly electromagnetism and matter, 592 pp,'' 1963.

\bibitem{Afek:2021bua}
G.~Afek, F.~Monteiro, B.~Siegel, J.~Wang, S.~Dickson, J.~Recoaro, M.~Watts, and D.~C. Moore, ``{Control and measurement of electric dipole moments in levitated optomechanics},'' {\em Phys. Rev. A}, vol.~104, no.~5, p.~053512, 2021.

\bibitem{jackson_classical_1999}
J.~D. Jackson, {\em Classical electrodynamics}.
\newblock New York, {NY}: Wiley, 3rd ed.~ed., 1999.

\bibitem{haynes2014crc}
W.~M. Haynes, {\em CRC handbook of chemistry and physics}.
\newblock CRC press, 2014.

\bibitem{stateexpansion}
E.~Bonvin, L.~Devaud, M.~Rossi, A.~Militaru, L.~Dania, D.~S. Bykov, O.~Romero-Isart, T.~E. Northup, L.~Novotny, and M.~Frimmer, ``State expansion of a levitated nanoparticle in a dark harmonic potential,'' {\em Phys. Rev. Lett.}, vol.~132, p.~253602, Jun 2024.

\bibitem{Chiaverini:2003}
J.~Chiaverini, S.~J. Smullin, A.~A. Geraci, D.~M. Weld, and A.~Kapitulnik, ``New experimental constraints on non-newtonian forces below $100\text{ }\text{ }\ensuremath{\mu}\mathrm{m}$,'' {\em Phys. Rev. Lett.}, vol.~90, p.~151101, Apr 2003.

\bibitem{Fosbinder-Elkins_2022}
H.~Fosbinder-Elkins, Y.~Kim, J.~Dargert, M.~Harkness, A.~A. Geraci, E.~Levenson-Falk, S.~Mumford, A.~Fang, A.~Kapitulnik, A.~Matlashov, D.~Kim, Y.~Shin, Y.~K. Semertzidis, Y.-H. Lee, N.~Aggarwal, C.~Lohmeyer, A.~Reid, J.~Shortino, I.~Lee, J.~C. Long, C.-Y. Liu, and W.~Snow, ``A method for controlling the magnetic field near a superconducting boundary in the ariadne axion experiment,'' {\em Quantum Science and Technology}, vol.~7, p.~014002, jan 2022.

\bibitem{Hudson1971}
W.~Hudson and R.~Jirberg, ``Superconducting properties of niobium films,'' NASA Technical Note NASA TN D-6030, NASA, 1971.

\bibitem{montoya:22}
C.~Montoya, E.~Alejandro, W.~Eom, D.~Grass, N.~Clarisse, A.~Witherspoon, and A.~A. Geraci, ``Scanning force sensing at micrometer distances from a conductive surface with nanospheres in an optical lattice,'' {\em Appl. Opt.}, vol.~61, pp.~3486--3493, Apr 2022.

\bibitem{grinin:2025}
A.~Grinin, A.~Dana, M.~Nguyen, E.~Alejandro, and A.~A. Geraci, ``A method for optically trapping nanospheres at micron range from a tilted mirror,'' 2025.
\newblock \texttt{\href{https://arxiv.org/abs/2504.18389}{arXiv:2504.18389[physics.optics]}}.

\bibitem{Sanz:2006}
J.~Pérez-Ríos and A.~S. Sanz, ``How does a magnetic trap work?,'' {\em American Journal of Physics}, vol.~81, p.~836–843, Nov. 2013.

\bibitem{extavour}
M.~H. Extavour, ``Fermions and bosons on an atom chip,'' 2009.
\newblock \texttt{\href{https://tspace.library.utoronto.ca/bitstream/1807/19030/11/Extavour_Marcius_H_T_200911_PhD_thesis.pdf}{PhD Thesis}}.

\bibitem{Bennett:1996gf}
C.~H. Bennett, D.~P. DiVincenzo, J.~A. Smolin, and W.~K. Wootters, ``{Mixed state entanglement and quantum error correction},'' {\em Phys. Rev. A}, vol.~54, pp.~3824--3851, 1996.

\bibitem{Fein2019}
Y.~Y. Fein, P.~Geyer, P.~Zwick, F.~Kia{\l}ka, S.~Pedalino, M.~Mayor, S.~Gerlich, and M.~Arndt, ``Quantum superposition of molecules beyond 25 kda,'' {\em Nature Physics}, vol.~15, pp.~1242--1245, Dec 2019.

\bibitem{Marshman:2023nkh}
R.~J. Marshman, S.~Bose, A.~Geraci, and A.~Mazumdar, ``Entanglement of magnetically levitated massive schr\"odinger cat states by induced dipole interaction,'' {\em Phys. Rev. A}, vol.~109, p.~L030401, Mar 2024.

\bibitem{Moorthy:2025fnu}
S.~N. Moorthy, A.~Geraci, S.~Bose, and A.~Mazumdar, ``{Magnetic noise in macroscopic quantum spatial superpositions},'' {\em Phys. Rev. A}, vol.~112, no.~2, p.~022416, 2025.

\bibitem{Barker:2022mdz}
P.~F. Barker, S.~Bose, R.~J. Marshman, and A.~Mazumdar, ``{Entanglement based tomography to probe new macroscopic forces},'' {\em Phys. Rev. D}, vol.~106, no.~4, p.~L041901, 2022.

\bibitem{Chakraborty:2023kel}
S.~Chakraborty, A.~Mazumdar, and R.~Pradhan, ``{Distinguishing Jordan and Einstein frames in gravity through entanglement},'' {\em Phys. Rev. D}, vol.~108, no.~12, p.~L121505, 2023.

\bibitem{Romero_Isart_2017}
O.~Romero-Isart, ``Coherent inflation for large quantum superpositions of levitated microspheres,'' {\em New Journal of Physics}, vol.~19, p.~123029, Dec. 2017.

\bibitem{zhou2024spindependentforceinvertedharmonic}
R.~Zhou, Q.~Xiang, and A.~Mazumdar, ``Spin-dependent force and inverted harmonic potential for rapid creation of macroscopic quantum superpositions,'' 2024.
\newblock \texttt{\href{https://arxiv.org/abs/2408.11909}{arXiv:2408.11909[quant-ph]}}.

\bibitem{braccini2024exponentialexpansionmassiveschrodinger}
L.~Braccini, A.~Serafini, and S.~Bose, ``Exponential expansion of massive schr\"{o}dinger cats for sensing and entanglement,'' 2024.
\newblock \texttt{\href{https://arxiv.org/abs/2408.11930}{arXiv:22408.11930[quant-ph]}}.

\bibitem{Geraci:2010}
A.~A. Geraci, S.~B. Papp, and J.~Kitching, ``Short-range force detection using optically cooled levitated microspheres,'' {\em Phys. Rev. Lett.}, vol.~105, p.~101101, Aug 2010.

\bibitem{moore2021searching}
D.~C. Moore and A.~A. Geraci, ``Searching for new physics using optically levitated sensors,'' {\em Quantum Science and Technology}, vol.~6, no.~1, p.~014008, 2021.

\bibitem{GiudiceDimopoulos}
S.~{Dimopoulos} and G.~F. {Giudice}, ``{Macroscopic forces from supersymmetry},'' {\em Physics Letters B}, vol.~379, pp.~105--114, Feb. 1996.

\bibitem{Montero2023}
M.~Montero, C.~Vafa, and I.~Valenzuela, ``The dark dimension and the swampland,'' {\em Journal of High Energy Physics}, vol.~2023, p.~22, Feb 2023.

\bibitem{kapner2007}
D.~J. Kapner, T.~S. Cook, E.~G. Adelberger, J.~H. Gundlach, B.~R. Heckel, C.~D. Hoyle, and H.~E. Swanson, ``Tests of the gravitational inverse-square law below the dark-energy length scale,'' {\em Phys. Rev. Lett.}, vol.~98, p.~021101, Jan 2007.

\end{thebibliography}
\bibliographystyle{ieeetr}

\end{document}